\documentclass[structabstract]{aa}  
\usepackage{graphicx}
\usepackage{txfonts}
\usepackage{natbib}
\def\teff{\mbox{$T_{\rm eff}$}}
\def\ebv{\mbox{$E(4405-5495)$}}
\def\rv{\mbox{$R_{5495}$}}
\def\al{\mbox{$A(\lambda)$}}
\def\mum1{\mbox{$\mu$m$^{-1}$}}
\def\chir{\mbox{$\chi^2_{\rm red}$}}
\begin{document}
   \title{The VLT-FLAMES Tarantula Survey}
   \subtitle{XVI. The optical and NIR extinction laws in 30 Doradus and the photometric determination of the effective temperatures of OB stars}

%   \subtitle{I. Overviewing the $\kappa$-mechanism}

   \author{J. Ma{\'\i}z Apell{\'a}niz\inst{1}
          \and
	  C. J. Evans\inst{2}
	  \and
	  R. H. Barb\'a\inst{3}
	  \and
	  G. Gr\"afener\inst{4}
	  \and
	  J. M. Bestenlehner\inst{4}
	  \and
	  P. A. Crowther\inst{5}
	  \and
	  M. Garc{\'\i}a\inst{6}
	  \and
   \linebreak
	  A. Herrero\inst{7,8}
	  \and
	  H. Sana\inst{9}
	  \and
   S. Sim\'on-D{\'\i}az\inst{7,8}
	  \and
	  W. D. Taylor\inst{2}
	  \and
	  J. Th. van Loon\inst{10}
	  \and
	  J. S. Vink\inst{4}
	  \and
	  N. R. Walborn\inst{9}
          }

   \institute{Instituto de Astrof{\'\i}sica de Andaluc{\'\i}a-CSIC, Glorieta de la Astronom\'{\i}a s/n, E-18\,008 Granada, Spain \\
	      \email{jmaiz@iaa.es}
         \and
              UK Astronomy Technology Centre, Royal Observatory Edinburgh, Blackford Hill, Edinburgh, EH9 3HJ, UK \\
         \and
              Departamento de F\'{\i}sica, Universidad de La Serena, Av. Cisternas 1200 Norte, La Serena, Chile \\
         \and
              Armagh Observatory, College Hill, Armagh BT61 9DG, UK \\
         \and
              Department of Physics \& Astronomy, Hounsfield Road, University of Sheffield, S3 7RH, UK  \\
         \and
              Centro de Astrobiolog{\'\i}a, CSIC-INTA, Ctra. Torrej\'on a Ajalvir km 4, E-28\,850 Torrej\'on de Ardoz, Madrid, Spain \\
         \and
              Instituto de Astrof{\'\i}sica de Canarias, E-38\,200 La Laguna, Tenerife, Spain \\
         \and
              Departamento de Astrof{\'\i}sica, Universidad de La Laguna, E-38\,205 La Laguna, Tenerife, Spain \\
         \and
              Space Telescope Science Institute, 3700 San Martin Drive, Baltimore, MD 21\,218, USA \\
         \and
              Institute for the Environment, Physical Sciences and Applied Mathematics, Keele University, ST5 5BG, UK \\
             }

   \date{Received 16 Jan 2014; accepted 13 Feb 2014}

% \abstract{}{}{}{}{} 
% 5 {} token are mandatory
 
  \abstract
   {The commonly used extinction laws of Cardelli et al. (1989) have limitations that, among other issues, hamper the determination of the effective
   temperatures of O and early B stars from optical and NIR photometry.}
   {We aim to develop a new family of extinction laws for 30 Doradus, check their general applicability within that region and elsewhere, and apply them to test
    the feasibility of using optical and NIR photometry to determine the effective temperature of OB stars.}
   {We use spectroscopy and NIR photometry from the VLT-FLAMES Tarantula Survey and optical photometry from HST/WFC3 of 30 Doradus and we analyze 
    them with the software code CHORIZOS using different assumptions, such as the family of extinction laws.}
   {We derive a new family of optical and NIR extinction laws for 30 Doradus and confirm its applicability to extinguished Galactic O-type systems. 
    We conclude that by using the new extinction laws it is possible to measure the effective temperatures of OB stars with moderate uncertainties and only
    a small bias, at least up to $\ebv \sim 1.5$ mag.}
   {}


   \maketitle
%
%________________________________________________________________

\section{Introduction}

Astronomy is entering a time when massive photometric surveys allow us to obtain information about a very large number of
objects. Projects such as Gaia and the Large Synoptic Survey Telescope (LSST) will reinforce this trend in the next decade. 
The main goal of these surveys is to measure the intrinsic properties of these objects, such as the effective temperature,
luminosity, and metallicity of stars; the mass, age, and metallicity of stellar clusters; or the redshift and type of galaxies. 
These surveys include not only large numbers of targets, but also detailed calibration mechanisms that lead to (internal) precisions 
and (external) accuracies at the level of one hundredth of a magnitude. In other words, we have not only data in large quantities
but also with high quality in the form of random and systematic errors that are significantly lower than what was typical twenty
years ago. This high photometric quality is also extended to space missions such as the Hubble Space Telescope (HST) and is
due to the stability of the space environment and the resources devoted to ensure the uniformity of the data. 

Despite such high quality, there is (and always will be) one obstacle for the derivation of the intrinsic properties of
astronomical objects: extinction. Every observation has to be corrected for the presence of dust between the target and the observer
and that can be (and in many cases is) the main limitation. In the 1980s the great success of the International Ultraviolet Explorer (IUE) satellite prompted a revived 
interest in the subject of extinction that culminated with the groundbreaking work of \citet[hereafter CCM]{Cardetal89}. 
that paper provided for the first time a family of extinction laws that extended from the IR to the UV while simultaneously characterizing
the type of extinction with a single parameter, \rv\ (see \citealt{Maiz13b} for a discussion on the name and the precise nature of the 
parameter). These two characteristics made the CCM laws a resounding success and the paper one of the most cited in astronomy in the last 
quarter of a century. 

Despite their unquestioned relevance, different studies in the last two decades have revealed several issues with some aspects of the
CCM laws:

\begin{itemize}
 \item the use of band-integrated [$E(B-V)$ and $R_V$] quantities to define the amount and type of extinction instead of their 
       monochromatic equivalents (\ebv\, and \rv, respectively)\footnote{It cannot be emphasized enough that using $R_V$ to parameterize an extinction law
       is a serious mistake. $R_V\equiv A_V/E(B-V)$ depends not only on the extinction law but also on the amount of extinction and the input SED. The 
       reader is referred to Figure 3 of \citet{Maiz13b} to quantify the effect. The parameter called $R_V$ in CCM is not really that
       (in the sense that an extinction law with a given value of that parameter does not yield that value of $A_V/E(B-V)$ for an arbitrary amount 
       of extinction and an arbitrary SED), but a monochromatic value. In addition, this type of effect in broad-band photometry has been known at least 
       since \citet{Blan57} but appears to be overlooked by a significant fraction of the astronomical community.};
 \item the validity of a fixed extinction law in the NIR;
 \item the functional form used in the optical;
 \item the reality of the correlation between \rv\ and UV extinction;
 \item the applicability of the laws beyond the \ebv\ and \rv\ values of the sample used to derive them;
 \item the photometric calibration of the filters.
\end{itemize}

These issues are discussed in \citet{Maiz13b}, where the reader is referred for details, and they are the reasons
that prompted us to attempt an improvement of the CCM laws, concentrating on the correction for extinction for photometric data, their 
most commonly used application.

This paper is part of a series on the VLT-FLAMES Tarantula Survey. The reader is referred to the first paper, \citet{Evanetal11a},
for details on the project. Within the series, this paper on the optical and NIR extinction law in 30 Doradus and its application to the 
determination of effective temperatures (\teff) is part of a subseries on extinction and the ISM. The subseries started with the work of
\citet{vanLetal13} on diffuse interstellar bands and neutral sodium and will continue with another paper on the spatial distribution of
extinction in 30 Doradus (Ma{\'\i}z Apell\'aniz et al. in preparation).

We start by describing the spectroscopic and photometric data in this paper. We then perform different experiments with the data by
processing them with CHORIZOS \citep{Maiz04c}. The results are discussed and possible future work is described. The paper
ends with three appendixes on (a) the detailed changes introduced by the new laws, (b) CHORIZOS and the spectral energy distributions (SEDs) used for this 
paper, and (c) the extinction along a sightline with more than one type of dust.

\section{Data}

\subsection{Spectral types and effective temperatures}

Our sample was selected mostly from the VFTS O-star sample \citep{Walbetal14} with the addition of some O and B stars also observed with VFTS. The majority of
the targets were observed by VFTS using the Medusa--Giraffe mode of the Fibre Large Array Multi-Element Spectrograph (FLAMES) instrument \citep{Pasqetal02}. 
Each star was observed with the LR02 and LR03 settings of the Giraffe spectrograph, which provided coverage of $\lambda$3960-5071\,\AA\ (at $R\equiv\Delta\lambda/\lambda$ of 7000 to 8500). 
Some of the stars in the R136 region (identifiable by their VFTS numbers above 1000), the massive cluster at its core, were observed with the LR02 setting using the 
ARGUS--Giraffe mode\footnote{These stars are not included in \citet{Walbetal14}. Neither are the B stars in this paper.}, 
which gives a comparable wavelength coverage but at a greater resolving power ($R\sim 10\,500$). Comprehensive classifications of these data for the 
O-type stars were presented by \citet{Walbetal14}, which also took into account binary companions detected by multi-epoch observations with the LR02+LR03 settings (see 
\citealt{Sanaetal13b}). The B stars were classified for this paper\footnote{A future VFTS B-star classification paper will appear as Evans et al., we have verified that there is a good
agreement between the independently derived spectral classifications for the two works.}.

The O-type stars in this paper sample the region of the 30~Doradus nebula imaged by the HST/WFC3 data described below, except for the very central part of R136 due to its 
dense stellar crowding. In the original target selection the only strong restriction was a faint-magnitude cut ($V$\,$\le$\,17 mag) to ensure sufficient signal-to-noise in 
the spectra of each target. The lack of color cuts should ensure that (moderately) reddened O-type stars were included, i.e., we are not strongly biased toward 
sightlines with low extinction. The O-type census obtained by VFTS is moderately complete across a 20$'$ field (excluding the central 0\farcm33). For instance, in the 
course of their analysis of the feedback from hot, luminous stars in 30~Dor (which also includes early B-type objects), \citet{Doraetal13} estimated that the Medusa VFTS 
observations were 76\% complete.

The spectral types were transformed into effective temperatures (\teff) using the calibration of \citet{Martetal05a} shifted upwards by 1000 K to account for the 
metallicity difference between the Milky Way and the LMC \citep{Mokietal07a,Doraetal13}. The shift is consistent with an ongoing analysis in the VFTS collaboration using FASTWIND grids (see
\citealt{SabSetal14} for some first results) and the IACOB-Grid Based Automatic Tool (IACOB-GBAT, \citealt{SimDetal11d}). From now on, the \teff\ derived from the spectral types 
will be called spectroscopic temperatures.

As described in Appendix~\ref{CHORIZOSgrids}, LMC-metallicity TLUSTY models are used as the intrinsic (extinction-free) SEDs for a given spectroscopic temperature. Since TLUSTY does not 
include wind effects, we should check for possible biases in the intrinsic colors. For a subsample of 12 stars analyzed individually within the VFTS collaboration with CMFGEN models (which 
include wind effects, see Bestenlehner et al. in preparation), we have compared the TLUSTY and the CMFGEN SEDs to check for possible systematic intrinsic color differences and we have found 
that they are very small ($\sim 0.01$ mag) when comparing models of the same \teff. Some slightly larger ($\sim 0.03$ mag) color variations were found in individual fits but these can be 
ascribed to differences on the order of 1000-2000 K between our spectroscopic temperatures (derived from the spectral types) and the star-by-star values of \teff\ derived from the CMFGEN 
analysis. Such variations are random, not systematic, and expected given the natural range of \teff\ existent within a given spectral type. Therefore, the use of values of \teff\ derived from 
spectral types and TLUSTY SEDs (as opposed to the more costly alternative of deriving individual SEDs for all the stars in the sample with, e.g. CMFGEN) may introduce a small amount of noise 
but no biases.

\subsection{NIR photometry}

\begin{table}
\caption{Sample, spectroscopic effective temperatures, and results of experiment 3.}
\centerline{
\begin{tabular}{lccc}
\hline
Object    & \teff   & \ebv            & \rv           \\
          & (K)     &                 &               \\
\hline
VFTS 385  & 42\,900 & 0.313$\pm$0.010 & 4.30$\pm$0.20 \\
VFTS 410  & 36\,900 & 0.492$\pm$0.027 & 5.15$\pm$0.50 \\
VFTS 422  & 43\,400 & 0.511$\pm$0.010 & 4.86$\pm$0.13 \\
VFTS 432  & 34\,900 & 0.489$\pm$0.010 & 4.52$\pm$0.15 \\
VFTS 436  & 36\,900 & 0.296$\pm$0.011 & 4.65$\pm$0.32 \\
VFTS 440  & 38\,500 & 0.320$\pm$0.010 & 4.57$\pm$0.20 \\
VFTS 451  & 37\,900 & 0.586$\pm$0.017 & 6.44$\pm$0.36 \\
VFTS 460  & 36\,900 & 0.677$\pm$0.010 & 3.97$\pm$0.09 \\
VFTS 464  & 32\,300 & 0.646$\pm$0.031 & 6.72$\pm$0.46 \\
VFTS 465  & 41\,900 & 0.740$\pm$0.009 & 5.27$\pm$0.09 \\
VFTS 472  & 39\,900 & 0.534$\pm$0.010 & 3.72$\pm$0.12 \\
VFTS 484  & 38\,900 & 0.386$\pm$0.010 & 5.09$\pm$0.18 \\
VFTS 491  & 39\,900 & 0.563$\pm$0.010 & 3.95$\pm$0.11 \\
VFTS 493  & 33\,900 & 0.636$\pm$0.010 & 4.00$\pm$0.11 \\
VFTS 494  & 35\,900 & 0.612$\pm$0.010 & 3.97$\pm$0.11 \\
VFTS 498  & 32\,300 & 0.441$\pm$0.011 & 5.03$\pm$0.27 \\
VFTS 505  & 32\,300 & 0.323$\pm$0.017 & 3.94$\pm$0.41 \\
VFTS 506  & 47\,700 & 0.324$\pm$0.010 & 4.25$\pm$0.18 \\
VFTS 508  & 32\,300 & 0.438$\pm$0.011 & 4.26$\pm$0.17 \\
VFTS 511  & 41\,900 & 0.402$\pm$0.010 & 4.16$\pm$0.17 \\
VFTS 512  & 47\,700 & 0.458$\pm$0.010 & 4.41$\pm$0.13 \\
VFTS 518  & 44\,300 & 0.529$\pm$0.010 & 4.00$\pm$0.12 \\
VFTS 520  & 26\,400 & 0.379$\pm$0.010 & 3.18$\pm$0.21 \\
VFTS 521  & 33\,900 & 0.366$\pm$0.012 & 4.84$\pm$0.35 \\
VFTS 525  & 26\,000 & 0.325$\pm$0.010 & 4.81$\pm$0.20 \\
VFTS 532  & 45\,900 & 0.455$\pm$0.010 & 4.13$\pm$0.14 \\
VFTS 543  & 33\,300 & 0.329$\pm$0.011 & 3.17$\pm$0.22 \\
VFTS 559  & 31\,100 & 0.367$\pm$0.011 & 4.72$\pm$0.28 \\
VFTS 560  & 32\,300 & 0.328$\pm$0.010 & 4.55$\pm$0.24 \\
VFTS 561  & 33\,900 & 0.376$\pm$0.010 & 4.17$\pm$0.20 \\
VFTS 563  & 30\,100 & 0.404$\pm$0.010 & 4.02$\pm$0.16 \\
VFTS 565  & 32\,300 & 0.298$\pm$0.011 & 4.67$\pm$0.33 \\
VFTS 566  & 45\,500 & 0.285$\pm$0.010 & 5.26$\pm$0.26 \\
VFTS 575  & 25\,000 & 0.257$\pm$0.010 & 3.14$\pm$0.29 \\
VFTS 577  & 39\,900 & 0.555$\pm$0.010 & 4.60$\pm$0.14 \\
VFTS 579  & 33\,900 & 0.376$\pm$0.034 & 6.33$\pm$0.83 \\
VFTS 585  & 37\,900 & 0.315$\pm$0.010 & 4.54$\pm$0.19 \\
VFTS 587  & 31\,100 & 0.279$\pm$0.010 & 4.43$\pm$0.28 \\
VFTS 591  & 25\,000 & 0.422$\pm$0.010 & 4.44$\pm$0.14 \\
VFTS 596  & 36\,900 & 0.396$\pm$0.010 & 4.06$\pm$0.16 \\
VFTS 597  & 34\,900 & 0.310$\pm$0.010 & 3.99$\pm$0.21 \\
\hline
\end{tabular}
}
\label{maintable}
\end{table}

\addtocounter{table}{-1}

\begin{table}
\caption{(continued).}
\centerline{
\begin{tabular}{lccc}
\hline
Object    & \teff   & \ebv            & \rv           \\
          & (K)     &                 &               \\
\hline
VFTS 598  & 29\,100 & 0.541$\pm$0.010 & 3.69$\pm$0.16 \\
VFTS 599  & 45\,500 & 0.340$\pm$0.010 & 4.51$\pm$0.18 \\
VFTS 601  & 40\,900 & 0.374$\pm$0.010 & 4.27$\pm$0.17 \\
VFTS 607  & 30\,100 & 0.297$\pm$0.010 & 5.04$\pm$0.31 \\
VFTS 608  & 43\,400 & 0.425$\pm$0.010 & 4.33$\pm$0.15 \\
VFTS 609  & 33\,300 & 0.371$\pm$0.012 & 4.15$\pm$0.39 \\
VFTS 611  & 35\,900 & 0.384$\pm$0.010 & 3.68$\pm$0.18 \\
VFTS 612  & 26\,400 & 0.413$\pm$0.010 & 4.06$\pm$0.17 \\
VFTS 616  & 27\,300 & 0.360$\pm$0.010 & 4.09$\pm$0.18 \\
VFTS 619  & 36\,900 & 0.372$\pm$0.010 & 4.28$\pm$0.18 \\
VFTS 635  & 31\,700 & 0.312$\pm$0.010 & 3.59$\pm$0.20 \\
VFTS 637  & 26\,400 & 0.292$\pm$0.010 & 3.09$\pm$0.25 \\
VFTS 646  & 28\,000 & 0.359$\pm$0.010 & 4.34$\pm$0.17 \\
VFTS 647  & 35\,900 & 0.339$\pm$0.012 & 4.62$\pm$0.50 \\
VFTS 648  & 40\,500 & 0.311$\pm$0.010 & 4.34$\pm$0.19 \\
VFTS 649  & 32\,300 & 0.327$\pm$0.010 & 3.93$\pm$0.20 \\
VFTS 651  & 37\,900 & 0.346$\pm$0.010 & 4.02$\pm$0.17 \\
VFTS 654  & 33\,900 & 0.305$\pm$0.010 & 4.90$\pm$0.24 \\
VFTS 656  & 36\,000 & 0.290$\pm$0.010 & 4.37$\pm$0.22 \\
VFTS 660  & 32\,300 & 0.280$\pm$0.010 & 4.99$\pm$0.27 \\
VFTS 664  & 36\,500 & 0.399$\pm$0.010 & 4.19$\pm$0.15 \\
VFTS 667  & 39\,900 & 0.341$\pm$0.010 & 3.99$\pm$0.18 \\
VFTS 676  & 26\,400 & 0.297$\pm$0.012 & 5.50$\pm$0.56 \\
VFTS 681  & 26\,400 & 0.286$\pm$0.010 & 3.80$\pm$0.24 \\
VFTS 686  & 25\,000 & 0.383$\pm$0.010 & 4.17$\pm$0.16 \\
VFTS 688  & 30\,100 & 0.375$\pm$0.010 & 3.65$\pm$0.16 \\
VFTS 692  & 28\,600 & 0.226$\pm$0.011 & 6.11$\pm$0.45 \\
VFTS 702  & 35\,900 & 0.583$\pm$0.011 & 4.30$\pm$0.17 \\
VFTS 705  & 26\,400 & 0.323$\pm$0.010 & 3.95$\pm$0.20 \\
VFTS 706  & 38\,900 & 0.431$\pm$0.010 & 3.54$\pm$0.13 \\
VFTS 707  & 27\,300 & 0.333$\pm$0.010 & 3.84$\pm$0.19 \\
VFTS 710  & 31\,700 & 0.287$\pm$0.010 & 3.16$\pm$0.22 \\
VFTS 712  & 26\,400 & 0.456$\pm$0.012 & 3.46$\pm$0.14 \\
VFTS 717  & 33\,300 & 0.466$\pm$0.010 & 4.00$\pm$0.14 \\
VFTS 728  & 29\,000 & 0.309$\pm$0.010 & 3.90$\pm$0.21 \\
VFTS 1002 & 31\,700 & 0.354$\pm$0.012 & 4.63$\pm$0.50 \\
VFTS 1006 & 38\,500 & 0.523$\pm$0.012 & 3.52$\pm$0.18 \\
VFTS 1007 & 38\,500 & 0.408$\pm$0.011 & 4.58$\pm$0.25 \\
VFTS 1018 & 41\,300 & 0.444$\pm$0.010 & 4.75$\pm$0.15 \\
VFTS 1020 & 44\,900 & 0.404$\pm$0.010 & 4.65$\pm$0.19 \\
VFTS 1028 & 43\,600 & 0.298$\pm$0.011 & 5.33$\pm$0.31 \\
VFTS 1035 & 33\,900 & 0.279$\pm$0.011 & 4.56$\pm$0.38 \\
\hline
\end{tabular}
}
\end{table}

Near-IR $JHK$ photometry for the vast majority of the VFTS targets is available from the Magellanic Clouds survey by \citet{Katoetal07} using the InfraRed Survey 
Facility (IRSF) 1.4~m telescope in South Africa. The IRSF ($JHK_{\rm s}$) magnitudes and their associated errors used in our analysis were presented in Table~6 in 
\citet{Evanetal11a}. The IRSF photometry was converted into the 2MASS system \citep{Skruetal06} by selecting a number of isolated stars with good S/N in both catalogs in 
the 30 Doradus region and deriving the corresponding linear transformations. We obtain the following relationships:

\begin{equation}
J_{\rm 2MASS} = J_{\rm IRSF} - 0.050 - 0.083(J_{\rm IRSF} - H_{\rm IRSF}), % Extra term is too small
\end{equation}

\begin{equation}
H_{\rm 2MASS} = H_{\rm IRSF} - 0.026 - 0.002(J_{\rm IRSF} - H_{\rm IRSF}), % Extra term is too small
\end{equation}

\begin{equation}
K_{\rm 2MASS} = K_{\rm IRSF} - 0.009 + 0.000(J_{\rm IRSF} - K_{\rm IRSF}). % Extra term is too small
\end{equation}

In order to ensure the accuracy of the NIR photometry, we compared the values with those of the VISTA VMC survey \citep{Cionetal11,Rubeetal12} and, for fields with 
strong nebulosity, with the ground-based photometry of \citet{Rubietal98a} and the HST photometry of \citet{Walbetal99a}. In cases with significant discrepancies we 
adjusted the values and increased the uncertainties of the used photometry.

\subsection{Optical photometry}

The optical photometry used in this paper was obtained with the UVIS channel of the Wide Field Camera 3 aboard HST as part of its Early Release Science program. Five
filters were used: F336W ($U$), F438W ($B$), F555W ($V$), F656N (H$\alpha$), and F814W ($I$), see \citet{DeMaetal11a} for details. Particular care was taken during the
preparation of the observations to include a wide range of exposure times in each filter; if that had not been done, the bright stars in the central region of 30 Doradus
(many of them included in the sample of this paper) would have been saturated. The multiple WFC3 frames were combined into a single image per filter using 
Multidrizzle\footnote{\tt http://stsdas.stsci.edu/multidrizzle/ .} after manually measuring the individual shifts and carefully selecting for each output pixel the information 
from the frames with the best S/N that were free of defects and not saturated. 

In principle, one can use PSF fitting to obtain the photometry of HST images. However, there are two reasons that advise against it in our case. (a) We need to combine 
the optical HST photometry with the ground-based NIR photometry: what appears as a single source in the latter may be (and in a number of cases is) multiple. (b) In some
regions of 30 Doradus the background is highly variable and a straightforward PSF fitting may underestimate the uncertainties associated with its subtraction. Hence, we
decided to do source-by-source aperture photometry in which we considered:

\begin{itemize}
 \item The number of point sources within the equivalent ground-based aperture. For isolated sources we used a single, small aperture. For multiple sources, we increased
       the aperture and applied an aperture correction or used multiple apertures (but see below for the final selection).
 \item Background subtraction. We considered a worst-case scenario in which the variations in the nearby background were taken as systematic instead of random. That is, 
       the value and the dispersion of the background were measured and the effect of the dispersion was added as an additional source of uncertainty by allowing the
       background to move up and down systematically (and not randomly on each pixel of the aperture). 
\end{itemize}

Considering all the uncertainty sources, a threshold of 0.02~magnitudes was determined to be needed for the uncertainty of all of the measured WFC3 magnitudes. We obtained 
the photometry for the five filters but we excluded F656N from the fit (since some O stars have H$\alpha$ filled or even in emission).

\subsection{Final sample selection}

Our initial sample consisted of 141 stars observed in the WFC3-ERS program. We analyzed each case individually and we eliminated those cases where [a] multiplicity was 
likely to yield a heterogeneous SED, [b] the photometry was internally or externally inconsistent (this could be caused by instrumental effects, e.g. undetected cosmic 
rays, or physical reasons, e.g. the system is photometrically variable), [c] the spectral type was of poor quality, or [d] the spectroscopic \teff\ was lower than 
24\,000 K (in these cases the differences in magnitude in the Balmer jump between adjacent spectral types become too large for our purposes). After the elimination 
process, we were left with 83 stars (67 O stars and 16 B stars). They are listed in Table~\ref{maintable}.

\section{Experiments}

The experiments in this paper are performed with v3.2 of the bayesian code CHORIZOS \citep{Maiz04c}. The procedure consists of fitting the available $UBVIJHK$-like
photometry to a family of synthetic SED models allowing for different parameters to be left free or kept fixed and for different extinction laws to be used. This method allows the
amount and type of extinction (along with possibly other quantities such as \teff\ or luminosity class) to be simultaneously fitted. See example 2 in \citet{Maiz04c} 
and \citet{Maizetal07}. The method has been called ``extinction without standards'' by \citet{FitzMass05b}. See Appendix~\ref{CHORIZOSgrids} and \citet{Maiz13a} for 
information on the used LMC-metallicity SED grid.

\subsection{Experiment 1: CCM laws and fixed \teff}

\begin{figure}
\includegraphics[width=\linewidth]{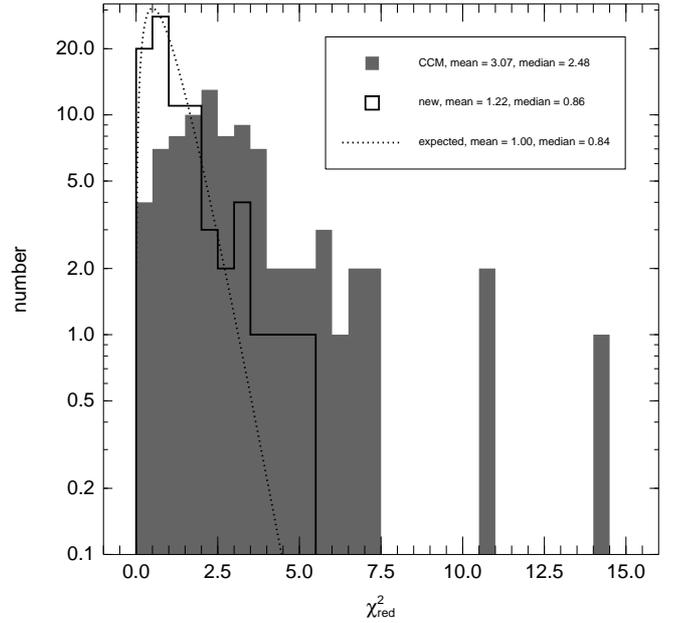}
\caption{\chir\ histograms for experiments 1 (filled) and 3 (continuous line). The dotted line shows the expected distribution for an ideal experiment. 
Note that the vertical scale is logarithmic.}
\label{chi2stats}
\end{figure}

%\begin{figure*}
%\centerline{\includegraphics[width=0.49\linewidth]{exp1_f336w_f438w.ps}}
%\centerline{\includegraphics[width=0.49\linewidth]{exp1_ebv_f438w.ps} \
%            \includegraphics[width=0.49\linewidth]{exp1_rv_f438w.ps}}
%\caption{Star-by-star residuals (observed minus best model) for [top] F438W vs. F336W, [bottom left] F438W vs. \ebv, and [bottom right] F438W vs. \rv\ in experiment 1. 
%See Figure~\ref{exp3} for the integrated residuals.}
%\label{exp1}
%\end{figure*}

\begin{figure}
\centerline{\includegraphics[width=0.92\linewidth, bb=28 40 566 550]{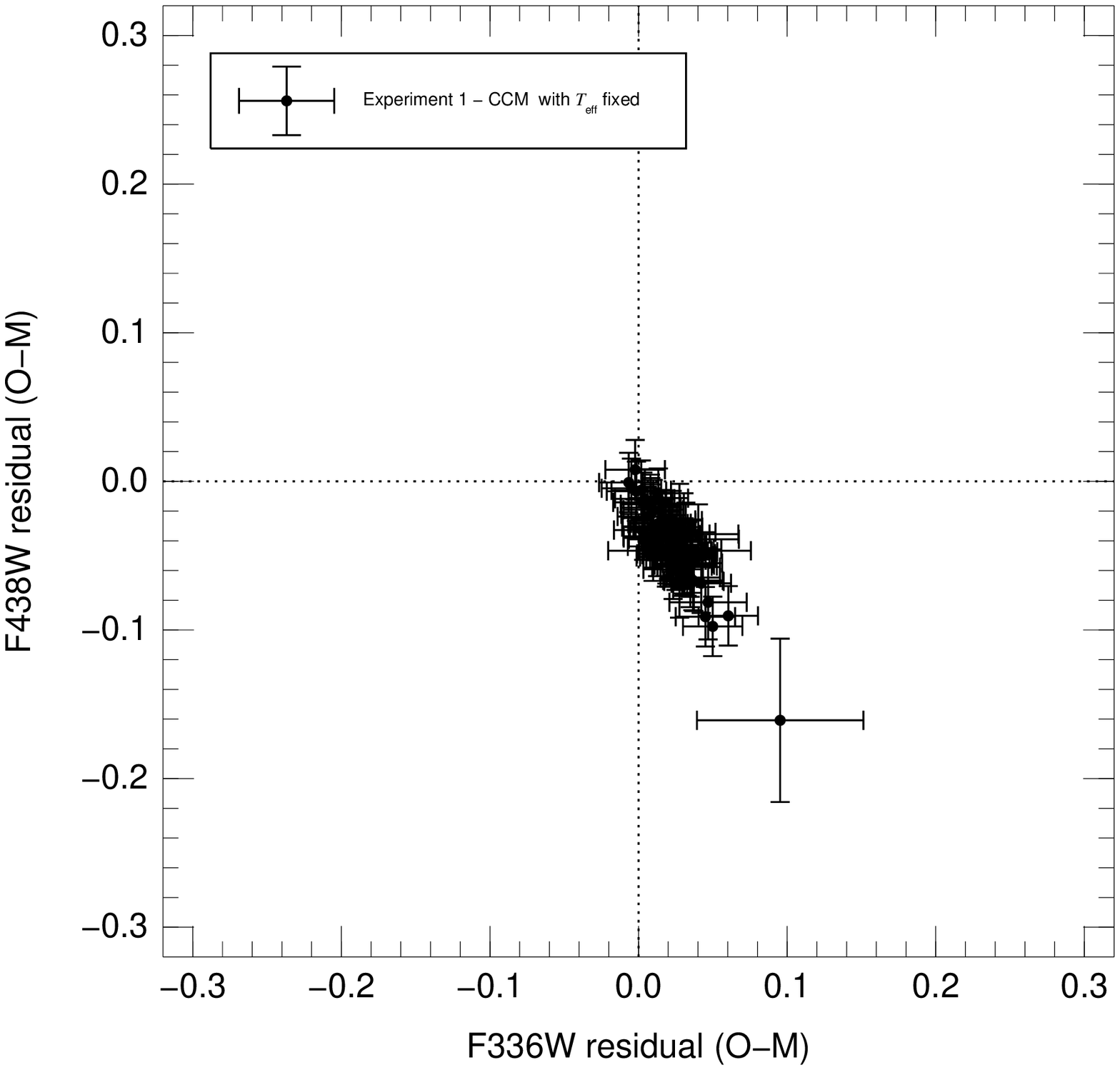}}
\centerline{\includegraphics[width=0.92\linewidth, bb=28 40 566 550]{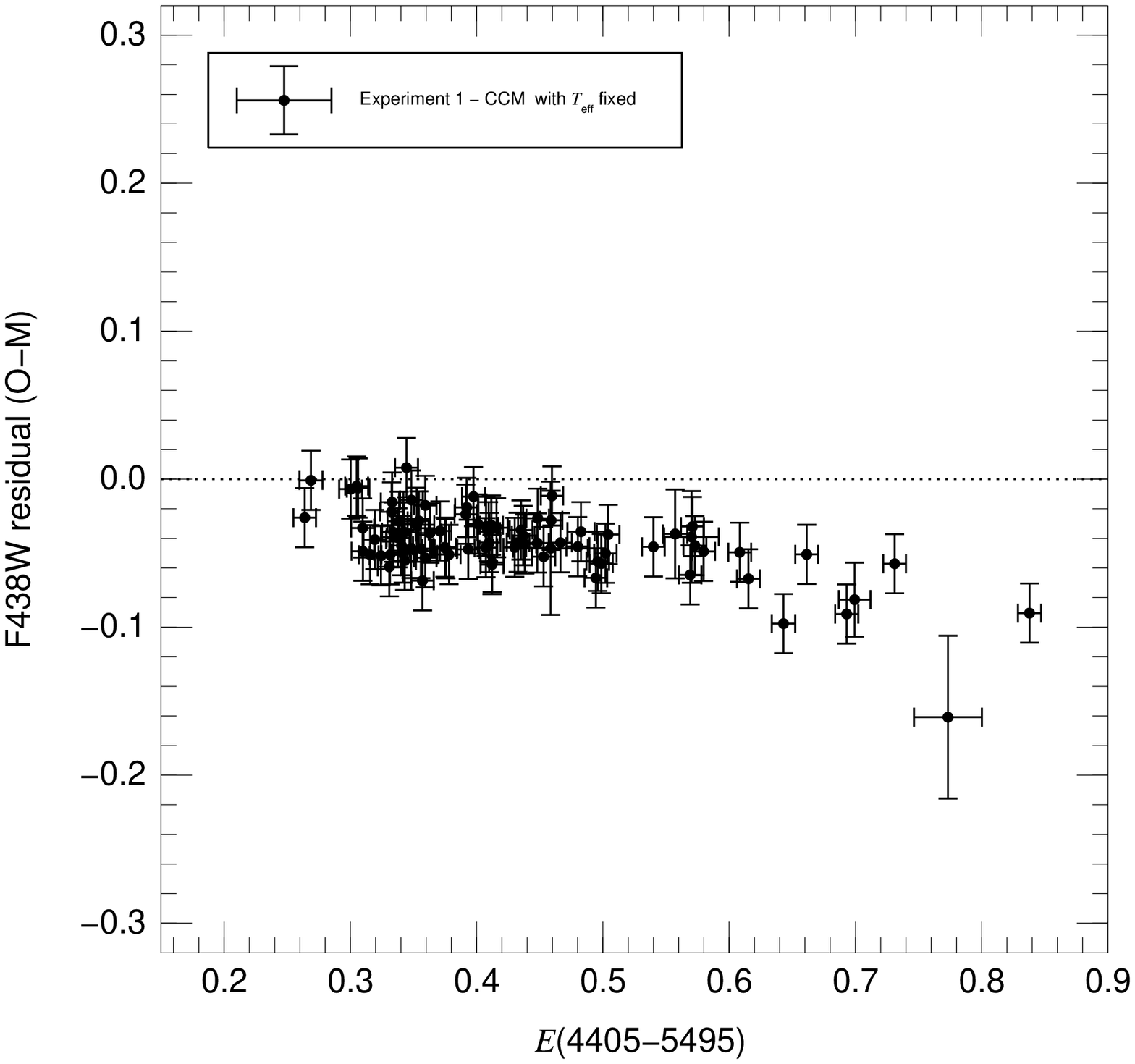}}
\centerline{\includegraphics[width=0.92\linewidth, bb=28 40 566 550]{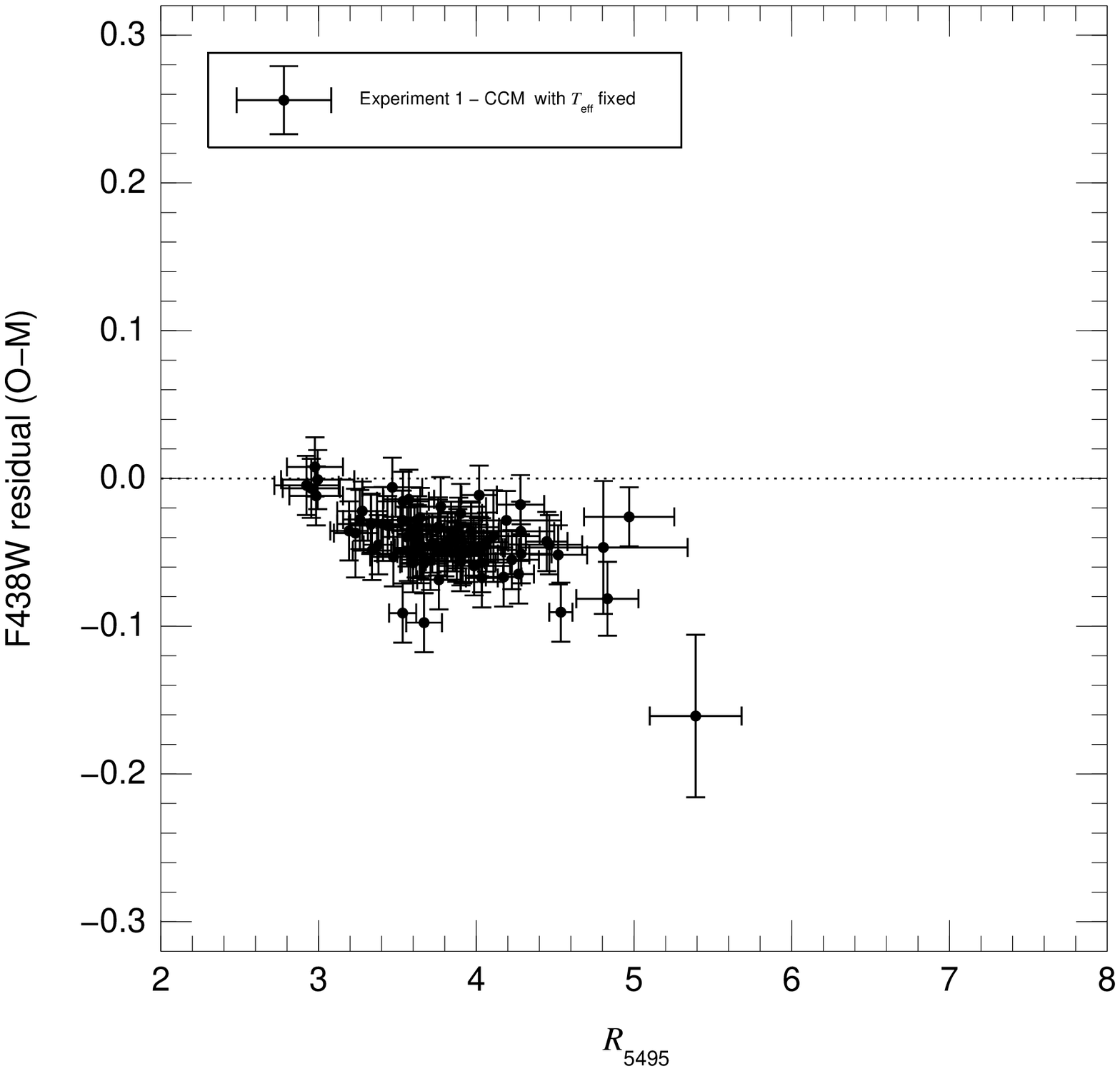}}
\caption{Star-by-star residuals (observed minus best model) for [top] F438W vs. F336W, [middle] F438W vs. \ebv, and [bottom] F438W vs. 
\rv\ in experiment 1. See Fig.~\ref{exp3} for the integrated residuals.}
\label{exp1}
\end{figure}

For our first experiment, we:

\begin{itemize}
 \item Fix the metallicity \citep{LanzHube03} and distance ($10^{4.7}$ pc) to the LMC values.
 \item Fix the \teff\ for each star to the value determined from its spectral type (see above).
 \item Use the CCM family of extinction laws.
 \item Leave three free parameters: (photometric) luminosity class (see Appendix~\ref{CHORIZOSgrids}), type of extinction [\rv], and amount of extinction [\ebv].
 \item Since for each star we have $M = 7$ magnitudes and $N=3$ free parameters, the solution has 4 degrees of freedom. 
\end{itemize}

Running CHORIZOS under these conditions yields results for \rv\ and \ebv\ with relatively good precision (small uncertainties) but the accuracy of the fit is unsatisfactory
due to the poor results of the \chir\ distribution (filled histogram in Figure~\ref{chi2stats}). The distribution has a mean of 3.07 and a median of 2.48 and its overall
appearance is different from the expected distribution (shown as a dotted line, mean of 1.00 and median of 0.84). The main peak is shifted towards the right and a significant tail 
is seen to have $\chir > 4.0$. This is a first sign that there is something wrong with either the photometric data, the range of parameters, the input SEDs, or the extinction laws. 
In order to find the cause we have to analyze the results in more detail.

Figure~\ref{exp1} shows some fit residuals (observed minus model) plotted against one another and against \ebv\ and \rv. Several effects are seen:

\begin{figure*}
\centerline{\includegraphics[width=0.49\linewidth]{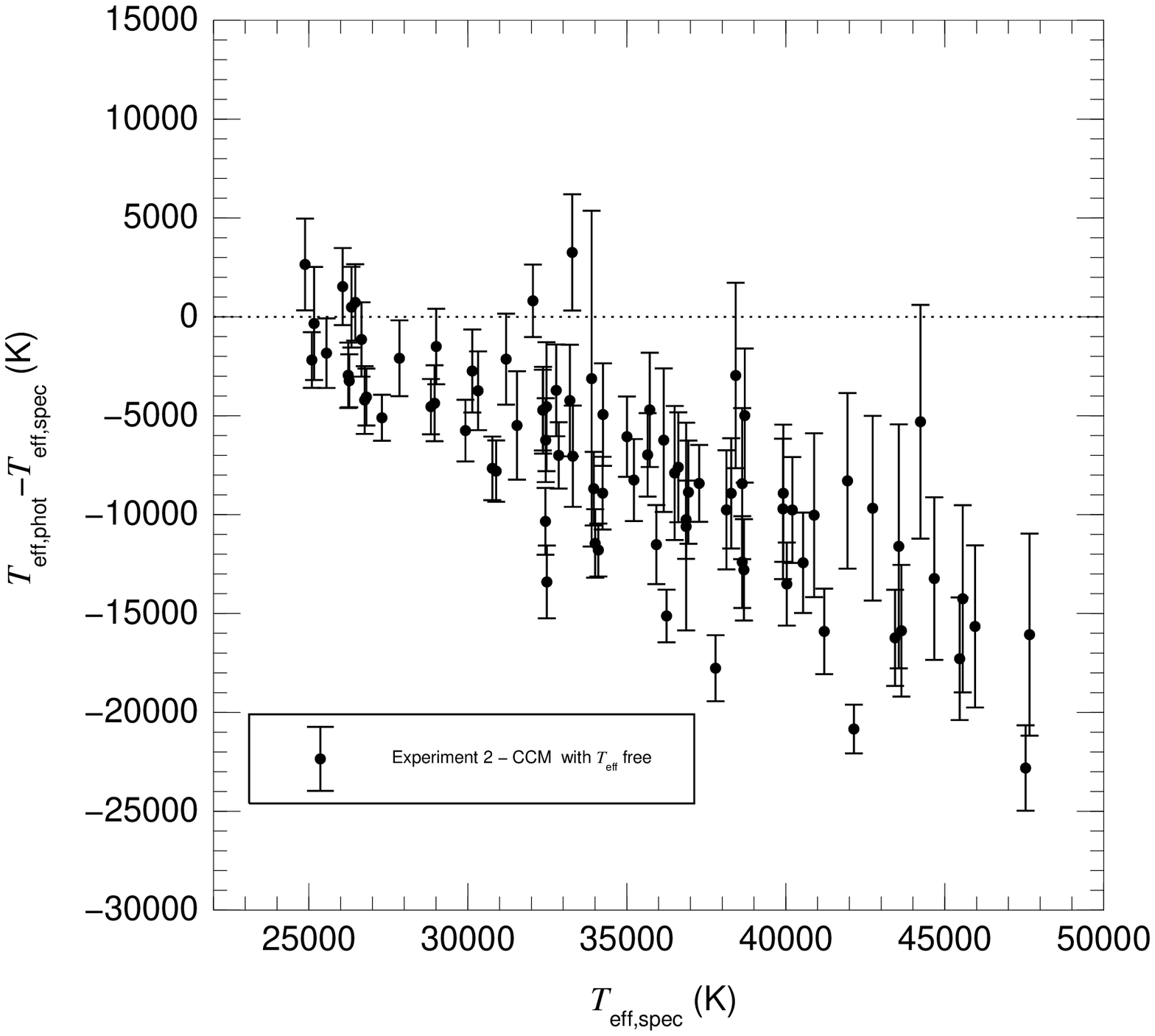} \
            \includegraphics[width=0.49\linewidth]{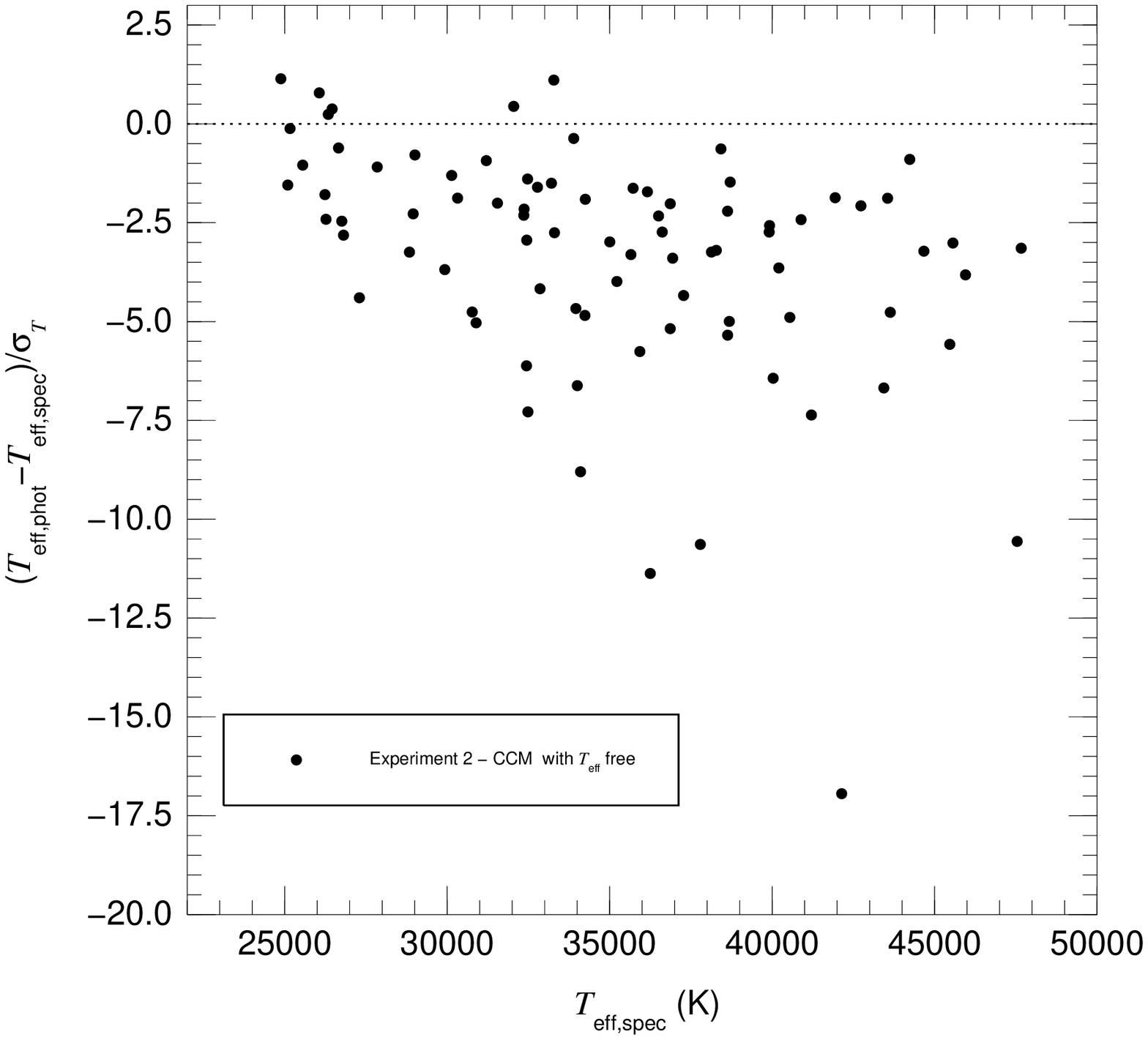}}
\caption{Results for experiment 2. The left panel shows the difference between the fitted \teff\ (derived from the photometry with CHORIZOS) and the \teff\
derived from the spectral classification (assumed to be the real \teff) as a function of the latter. The right panel is the same plot but with the vertical axis
normalized by the uncertainties (an ideal solution would have mean of zero and a standard deviation of 1 without depending on \teff). Note that a small amount of
random noise (standard deviation of 200 K) has been introduced in the horizontal values to decrease the superposition between different objects.}
\label{exp2}
\end{figure*}

\begin{itemize}
 \item The F336W ($U$) and F438W ($B$) residuals are strongly anticorrelated and these filters carry a good fraction of the weight of \chir\ in the targets where \chir\ is high. 
       CHORIZOS is finding as the best possible solution an intermediate SED that is too bright in $U$ and too dim in $B$ but cannot find an optimal solution because it is 
       not within the allowable ones. At a lower level, similar anticorrelations are found between other residuals from adjacent filters.
 \item The F438W residual is correlated with the amount of extinction. This points towards the extinction law as the culprit, since for low values of the extinction the
       residuals show only a small bias in their distribution.
 \item The F438W is also clearly correlated with \rv. For $\rv\sim 3$, the residual distribution is centered around zero but for high values there is a clear offset. 
       This indicates that the CCM laws for the lower values of \rv\ provide a better fit than for higher values.
\end{itemize}

Therefore, these results lead us to think that the problem is in the exact form of the CCM extinction laws, a hypothesis that we will test in subsequent experiments.
If the CCM family of extinction laws does not provide an adequate solution, one should first test an existing alternative. \citet{Fitz99} presented a family of \rv-dependent
extinction laws (from now on, we will call them F99 laws) which we have also implemented in CHORIZOS. Executing experiment 1 with these alternate laws we find that 
they do not provide an adequate solution, either. The \chir\ distribution has a mean of 4.43 and a median of 3.95, i.e. even worse than the CCM result. 

\subsection{Experiment 2: CCM laws and variable \teff}

Before attempting a modification of the CCM family of extinction laws, we perform a second experiment with them and an important modification on the conditions: we
leave \teff\ as a free parameter. This increases $N$ to 4 and leaves just 3 degrees of freedom. It also introduces a new measure of the accuracy of the fit: how do the
calculated (photometric) \teff\ compare with their spectroscopic counterparts.

The results of the second experiment look encouraging at first. They yield a \chir\ distribution with a mean of 1.09 and a median of 0.91, which are very similar to the
results expected for an ideal experiment with 3 degrees of freedom (1.00 and 0.78, respectively). The problem arises from the differences between the photometric and spectroscopic 
\teff. The photometric values are lower by an average of 7700 K and there is a significant trend that makes the results get worse for higher values of \teff\ (Figure~\ref{exp2}). 
Looking at the results in a different way, the spectroscopic temperature range of 30\,000-40\,000~K is approximately mapped into a 25\,000-30\,000~K range in photometric temperature (shifted 
downwards but also compressed). When normalized by their (CHORIZOS-derived photometric) uncertainties, the difference between the two values has a mean of $-3.33$ and a standard deviation of 
$2.94$, a long way from the expected respective values of $0.0$ and $1.0$ in an ideal experiment. Note that the intrinsic scatter in \teff\ within a given spectral type and luminosity class is 
1000-2000 K, much lower than the offset detected here.

This second experiment allows us to draw two important conclusions:

\begin{itemize}
 \item CCM laws cannot be used to accurately derive the \teff\ of O stars from photometry because they introduce a significant bias even for moderate values of \ebv\ (most of the 
       stars in our sample are in the range between 0.3 and 0.7 mag).
 \item The main difference in the $UBVIJHK$ photometry of e.g. a 30\,000 K and a 40\,000 K star of similar gravities lies in the $U-B$-like color because in the optical and NIR 
       their SEDs are relatively well described by the Rayleigh-Jeans law, with the main difference arising from the Balmer jump\footnote{Note that the WFC3 F336W-F438W color
       provides a cleaner measurement of the Balmer jump than Johnson's $U-B$, see \citet{Maiz05b,Maiz06a,Maiz07a}.}. Therefore, the fact that experiment 2 yields a good \chir\
       distribution but with the wrong temperatures is telling us that the problem of the CCM laws is in the same wavelength region (3000-5000 \AA), since it should be possible to
       tweak the laws in that region to counteract the artificial temperature shift by an equivalent change in the $U-B$-like colors. More specifically, the results of experiments 1
       and 2 indicate that the problem with the CCM laws appears to be concentrated in the $U$ band for high values of \rv.
\end{itemize}

\subsection{Experiment 3: New laws and fixed \teff}

%\begin{figure*}
%\centerline{\includegraphics[width=0.49\linewidth]{exp3_f336w_f438w_cont.ps}}
%\centerline{\includegraphics[width=0.49\linewidth]{exp3_ebv_f438w_cont.ps} \
%            \includegraphics[width=0.49\linewidth]{exp3_rv_f438w_cont.ps}}
%\caption{Integrated residuals (observed minus best model and calculated assuming bidimensional gaussian distributions for each individual point) for 
%[top] F438W vs. F336W, [bottom left] F438W vs. \ebv, and [bottom right] F438W vs. \rv\ in experiments 1 and 3. 
%The results for experiment 1 are shown as black and white contour plots and are the result of integrating the data in Figure~\ref{exp1} (note that a slightly smaller range of
%magnitudes is shown here). The results for experiment 3 are shown as color-filled density diagrams. The three small squares in the top panel mark the centers 
%of the distributions for experiments 1 and 3 and for the ideal solution. The dotted lines in the bottom panels are the linear fits to the residuals for experiments 1 and 3 and
%the ideal solution.}
%\label{exp3}
%\end{figure*}

\begin{figure}
\centerline{\includegraphics[width=0.80\linewidth, bb=28 40 566 550]{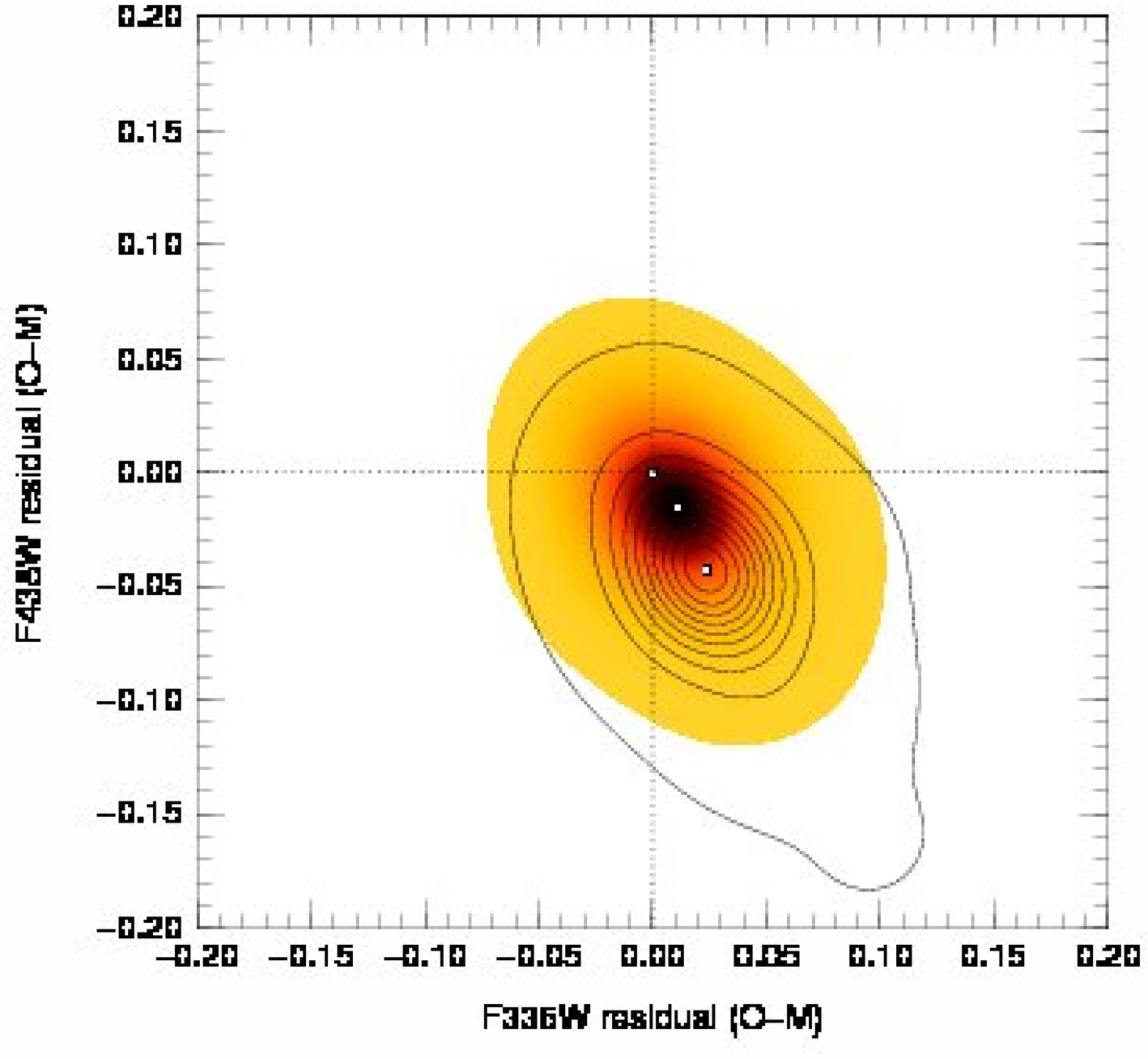}}
\centerline{\includegraphics[width=0.80\linewidth, bb=28 40 566 550]{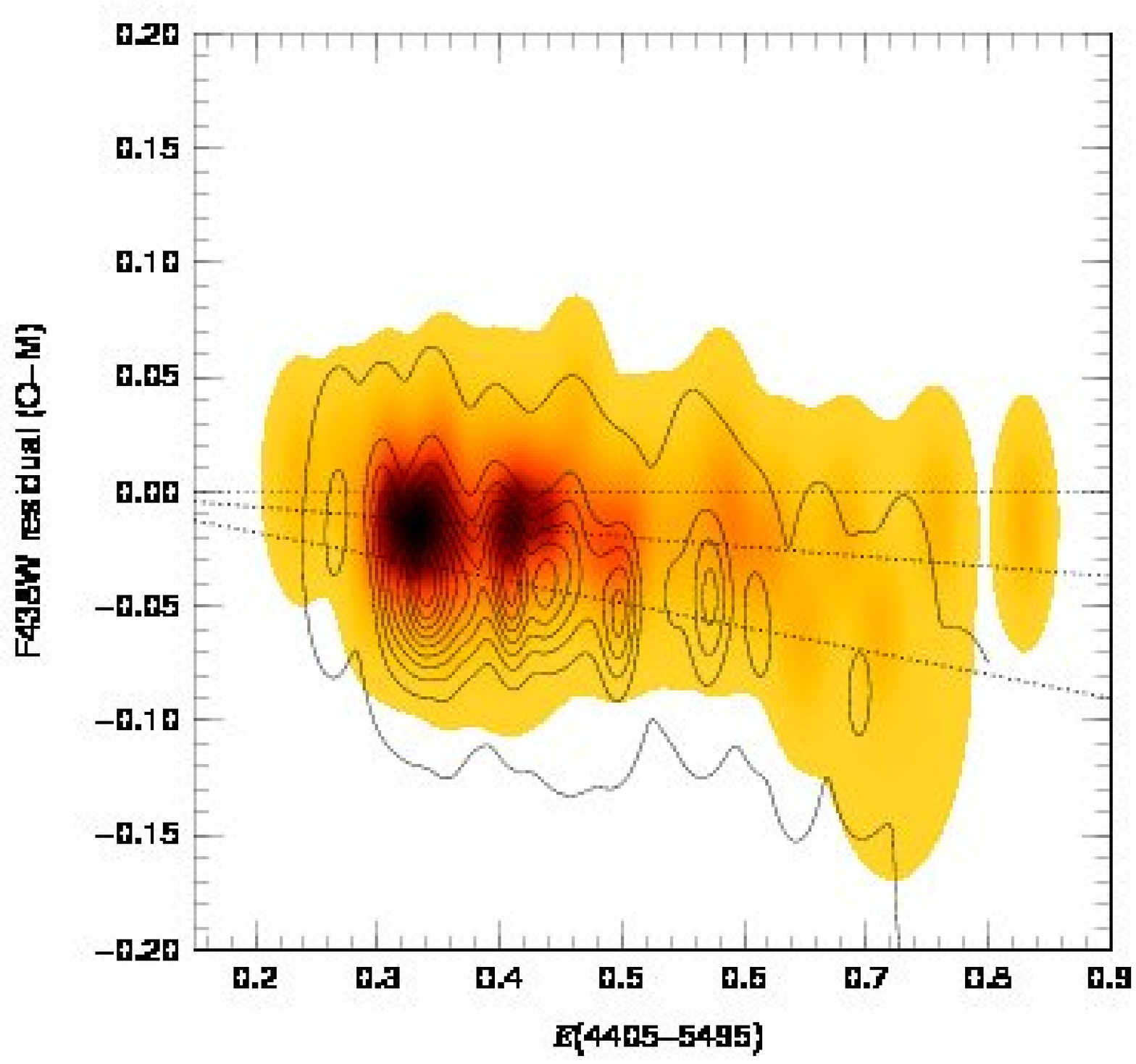}}
\centerline{\includegraphics[width=0.80\linewidth, bb=28 40 566 550]{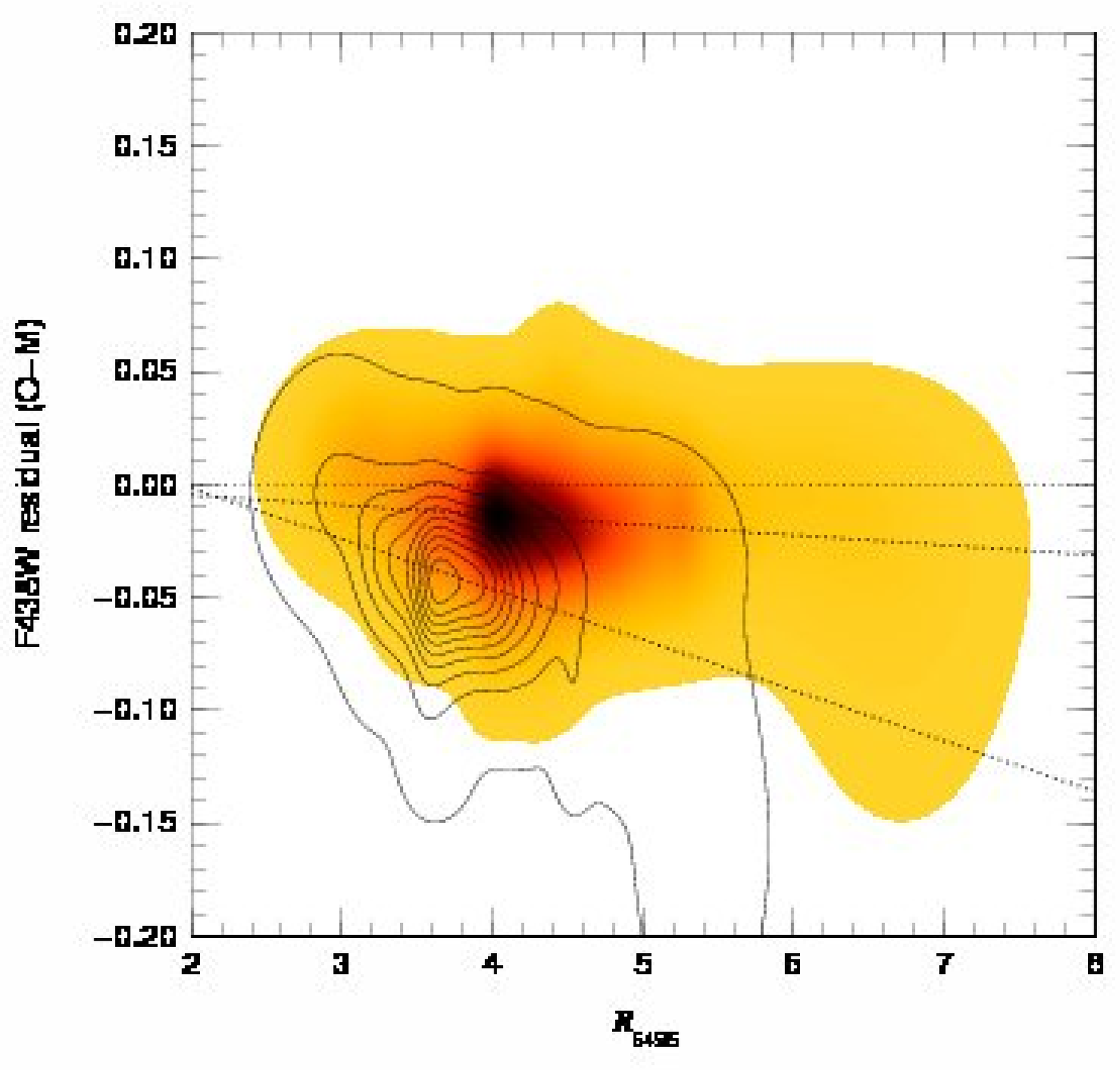}}
\caption{Integrated residuals (observed minus best model and calculated assuming bidimensional gaussian distributions for each individual 
point) for [top] F438W vs. F336W, [middle] F438W vs. \ebv, and [bottom] F438W vs. \rv\ in experiments 1 and 3. The results for experiment 1 
are shown as black and white contour plots and are the result of integrating the data in Figure~\ref{exp1} (note that a slightly smaller 
range of magnitudes is shown here). The results for experiment 3 are shown as color-filled density diagrams. The three small squares in the 
top panel mark the centers of the distributions for experiments 1 and 3 and for the ideal solution. The dotted lines in the bottom panels 
are the linear fits to the residuals for experiments 1 and 3 and the ideal solution.}
\label{exp3}
\end{figure}

\begin{figure*}
\centerline{\includegraphics[width=0.49\linewidth, bb=50 28 566 566]{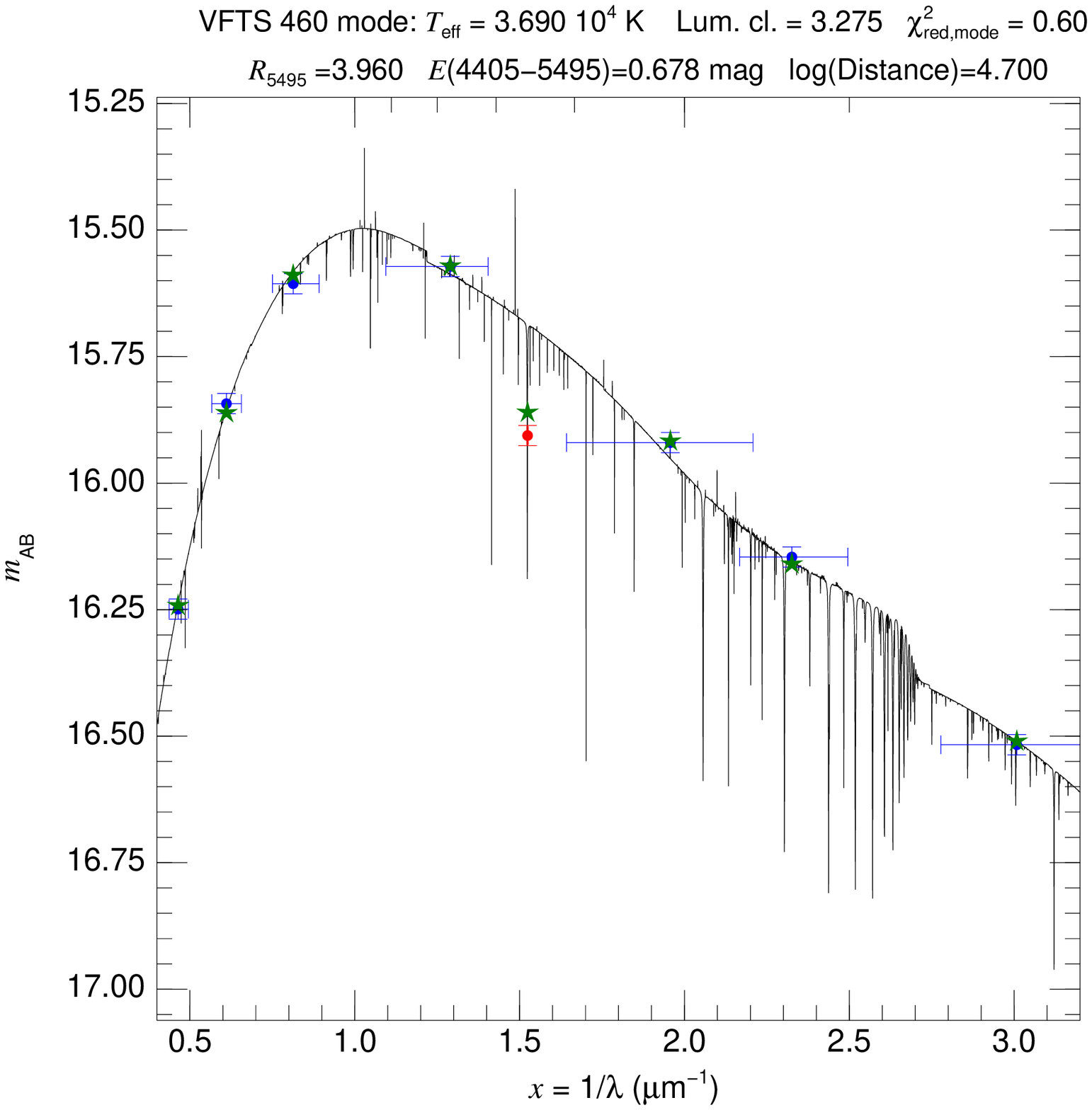} \
            \includegraphics[width=0.49\linewidth, bb=50 28 566 566]{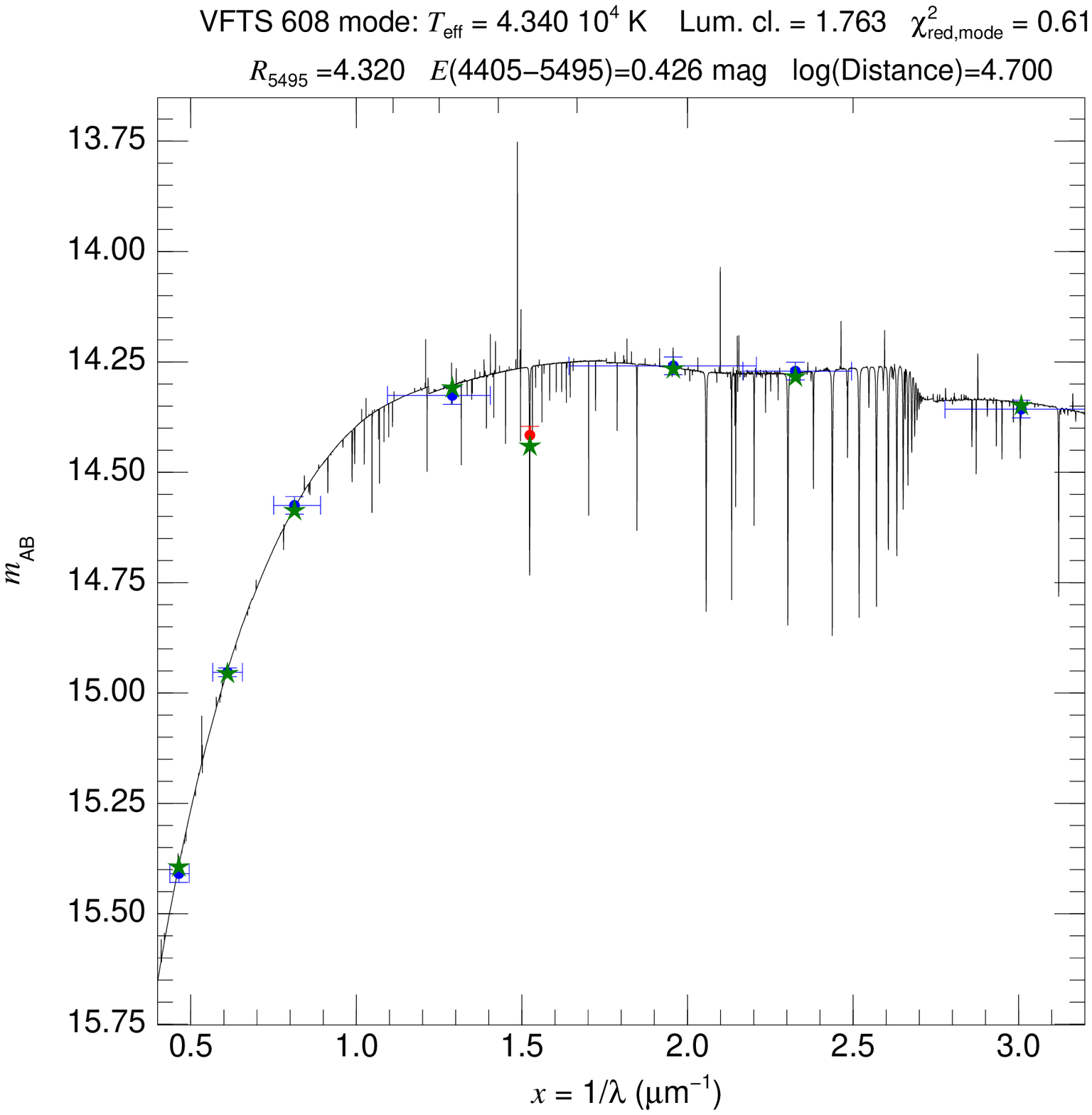}}
\centerline{\includegraphics[width=0.49\linewidth, bb=50 28 566 566]{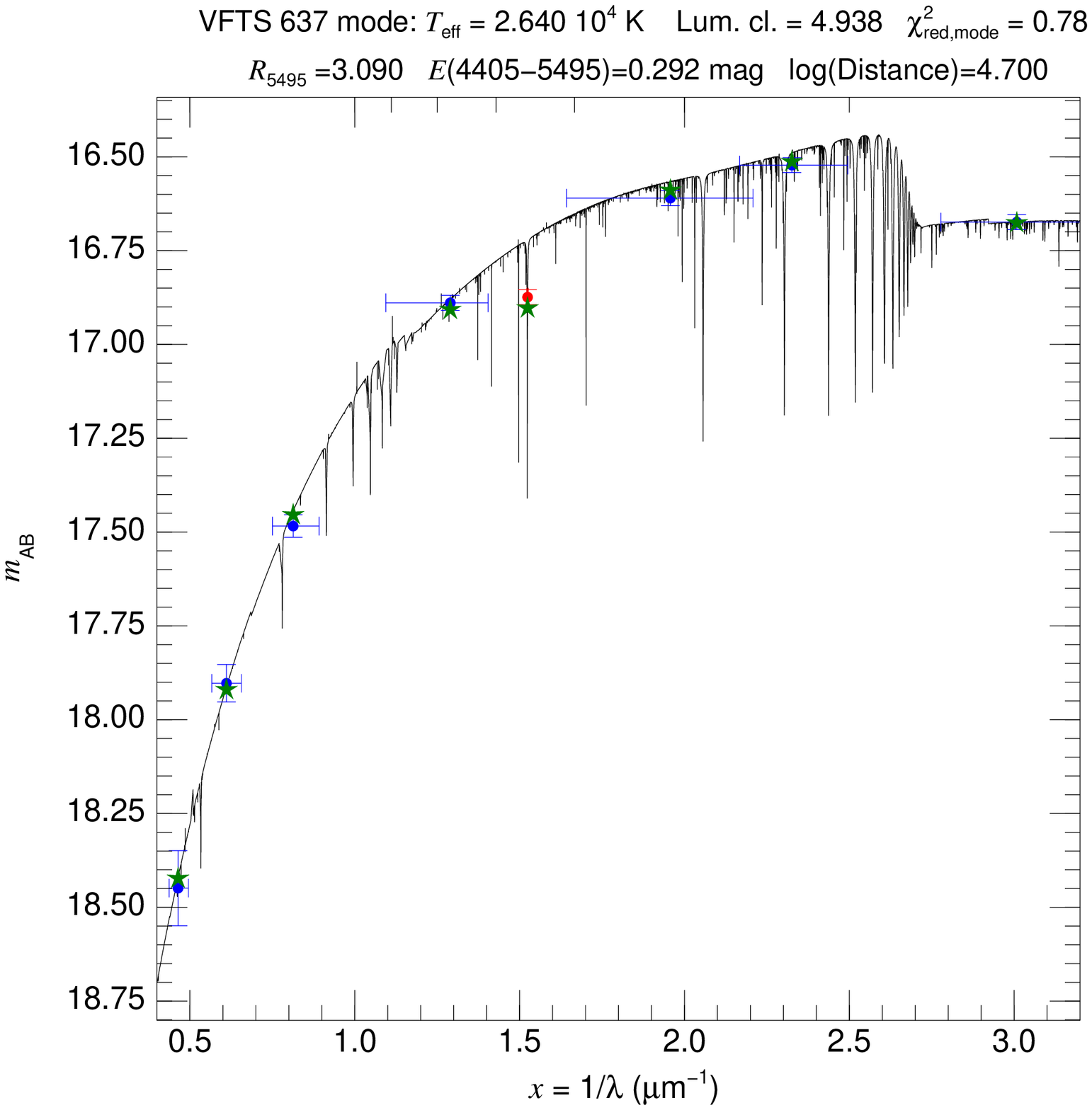} \
            \includegraphics[width=0.49\linewidth, bb=50 28 566 566]{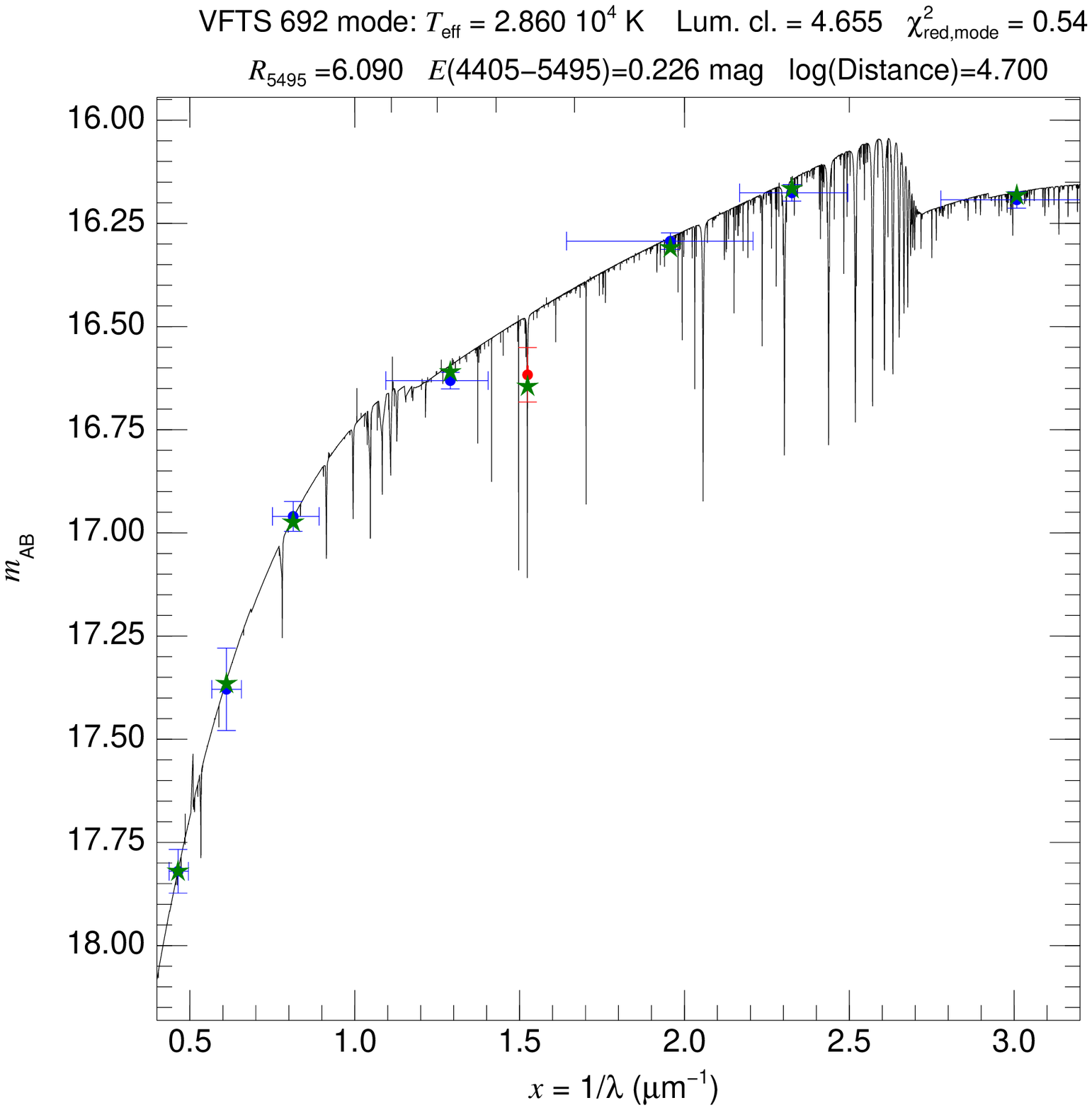}}
\caption{Four examples of results from experiment 3. The continuous line shows the best (mode) TLUSTY SED output from CHORIZOS and the green stars are the associated synthetic 
photometry (from right to left, F336W, F438W, F555W, F656N, F814W, $J$, $H$, and $K_{\rm s}$). The blue points with error bars (horizontal for indicative filter extent, vertical for 
uncertainty) show the input photometry used for the fit. The red points with error bars show the input F656N filter (not used for the fit). The vertical axes are in AB magnitudes
\citep{synphot}.}
\label{SED_exp3}
\end{figure*}

%\begin{figure*}
%\centerline{\includegraphics[width=0.49\linewidth, bb=50 28 566 566]{exp1_exp3_ebv.ps} \
%            \includegraphics[width=0.49\linewidth, bb=50 28 566 566]{exp1_exp3_rv.ps}}
%\centerline{\includegraphics[width=0.49\linewidth, bb=50 28 566 566]{exp1_exp3_af555w.ps}}
%\caption{Values obtained for \ebv\ (top left), \rv\ (top right), and $A_{\rm F555W}$ in the first and third experiments. The dotted line shows a 1:1 relationship.}
%\label{ebv_rv}
%\end{figure*}

\begin{figure}
\centerline{\includegraphics[width=0.90\linewidth, bb=28 28 566 550]{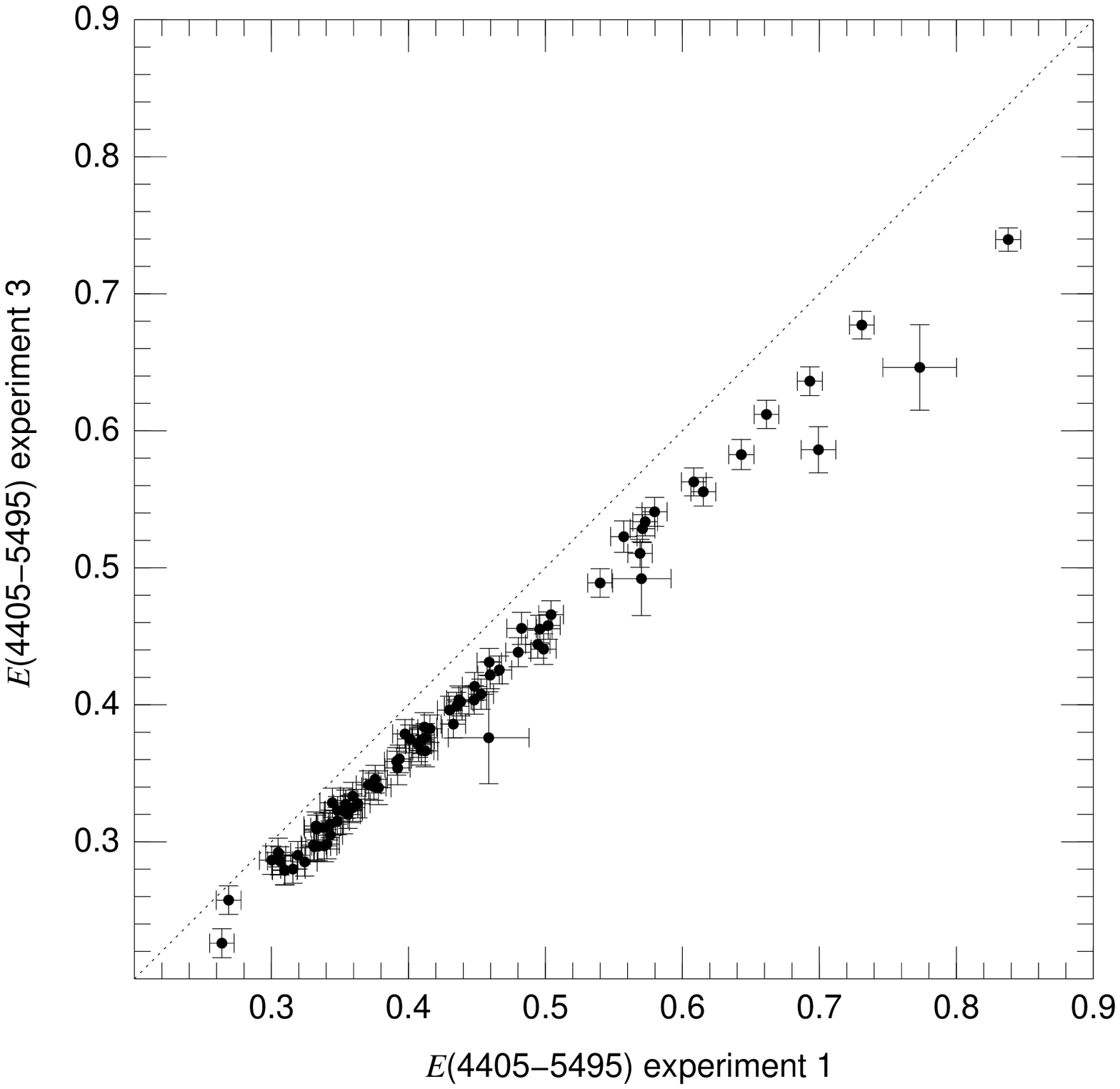}}
\centerline{\includegraphics[width=0.90\linewidth, bb=28 28 566 550]{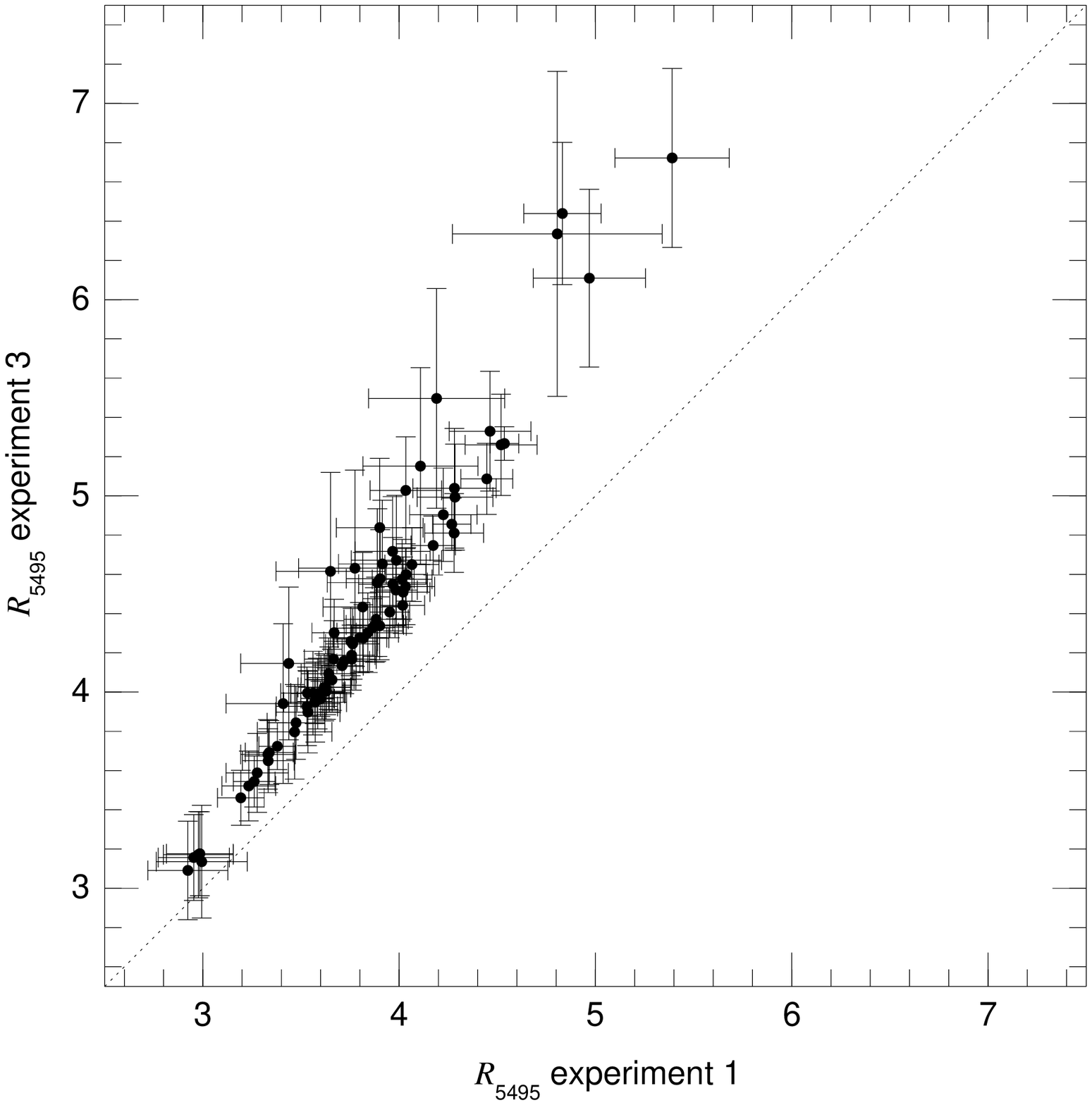}}
\centerline{\includegraphics[width=0.90\linewidth, bb=28 28 566 550]{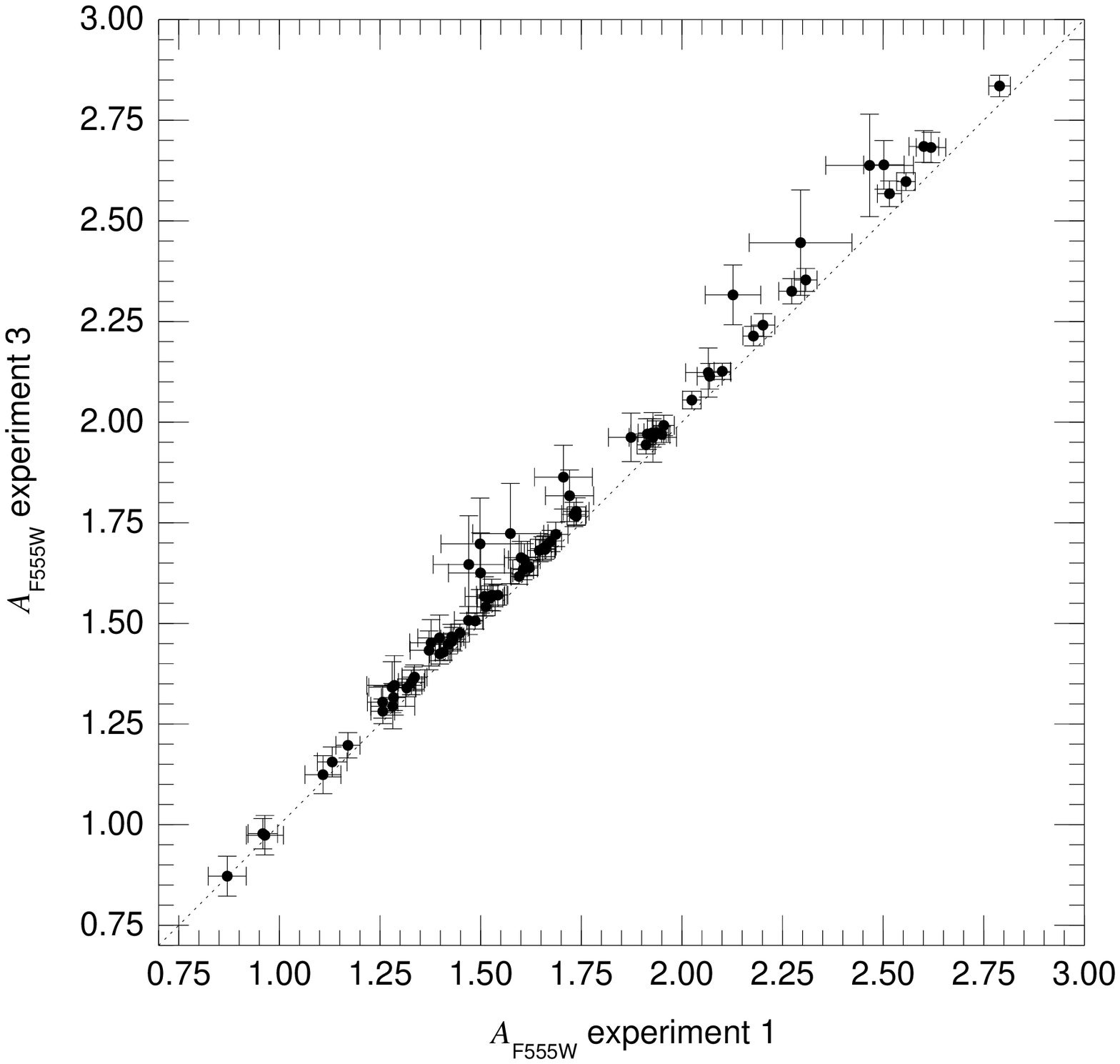}}
\caption{Values obtained for \ebv\ (top), \rv\ (middle), and $A_{\rm F555W}$ (bottom) in the first and third experiments. The dotted line 
shows a 1:1 relationship.}
\label{ebv_rv}
\end{figure}

\begin{figure*}
\centerline{\includegraphics[width=0.49\linewidth]{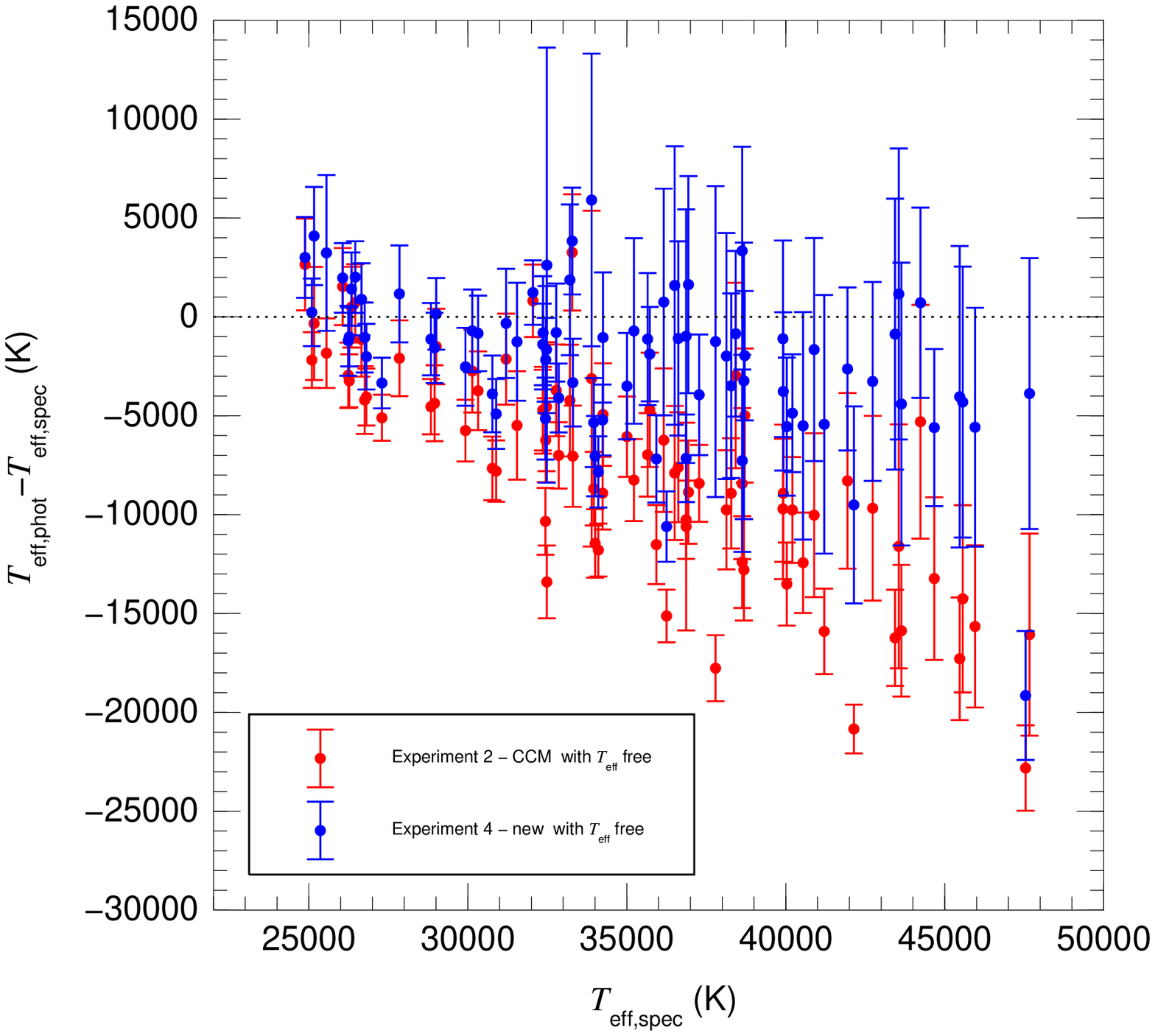} \
            \includegraphics[width=0.49\linewidth]{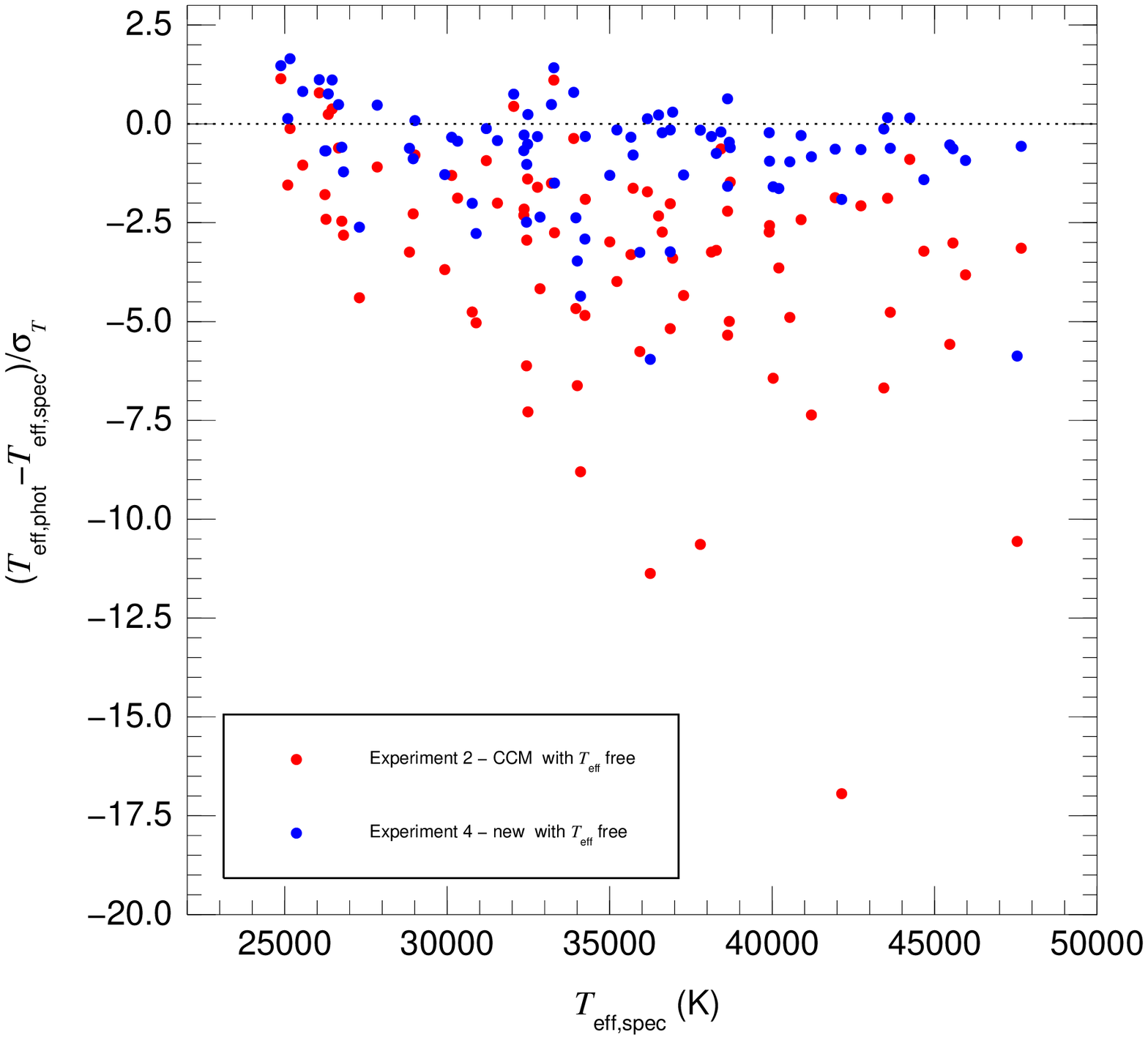}}
\caption{Results for experiments 2 (red) and 4 (blue). The left panel shows the difference between the fitted \teff\ (derived from the photometry with CHORIZOS) and the \teff\
derived from the spectral classification (assumed to be the real \teff) as a function of the latter. The right panel is the same plot but with the vertical axis
normalized by the uncertainties (an ideal solution would have mean of zero and a standard deviation of 1 without depending on \teff). Note that a small amount of
random noise (standard deviation of 200 K) has been introduced in the horizontal values to decrease the superposition between different objects.}
\label{exp4}
\end{figure*}

The two previous experiments allowed us to qualitatively determine in which direction we should introduce corrections to the CCM laws in order to provide a better fit to the
observed photometry (making the \chir\ distribution look more similar to the expected distribution in the first case and yielding better temperature estimates in the second).
The ideal way to quantitatively derive these corrections would be to obtain good-quality spectrophotometry with a large wavelength coverage (i.e. the same method \citealt{Whit58}
used) but, unfortunately, that is not available for an appropriate sample. We have to resort to an iterative process in which we estimate a new family of laws (using the qualitative
criteria above) and proceed by trial and error running CHORIZOS with the estimate, analyzing the behavior of the photometric residuals in the experiment 1 equivalent (from now on,
experiment 3) and the temperature differences in the experiment 2 equivalent (from now on, experiment 4) to improve on the results. After more than 20 iterations (each one taking 
several days), we arrived at the new family. Details of the procedure are given in Appendix~\ref{newlaws}. 

The improvements introduced by the final form of the new laws can be seen in the first place in Figure \ref{chi2stats}, which compares the results of experiments 1 and 3. The new 
laws yield a \chir\ distribution with a mean of 1.22 and a median of 0.86. The second value is very similar to the ideal result of 0.84 while the first one is only slightly higher
than the perfect result of 1.00. Most of the difference can be attributed to the existence of four stars with \chir\ between 3.5 and 5.5. Additional improvements in experiment 3
with respect to experiment 1 can be seen in Figure~\ref{exp3}. In the top panel, the anticorrelation between 
F336W and F438W has diminished considerably and the outliers have disappeared. In the bottom two panels we see that the F438W residuals now have a weaker dependence on \ebv\ and 
\rv, respectively. In summary, {\it the new laws provide a significantly better fit to the observed photometry if spectroscopic temperatures are used as an input to constrain the
unextinguished SED}.

Four examples of results from experiment 3 are shown in Figure~5. Note the small extent of the vertical error bars (photometric uncertainties) and the good agreement between them
and the synthetic photometry (green stars). Figure~\ref{ebv_rv} shows the relationship between the \ebv, \rv, and $A_{\rm F555W}$ results for experiments 1 and 3 (see also 
Table~\ref{maintable}). In experiment 3 the results for \ebv\ are consistently lower and the results for \rv\ consistently higher than those in experiment 1. Note, however, that the 
results for $A_{\rm F555W}$ remain almost unchanged. A linear regression yields $A_{{\rm F555W,exp}\,3} = -0.004+1.023 A_{{\rm F555W,exp}\,1}$, so that 
for e.g.  $A_{{\rm F555W,exp}\,1} = 2.00$~mag, $A_{{\rm F555W,exp}\,3}$ is typically $2.04$~mag. Note, also, that in many cases the relative errors in $A_{\rm F555W}$ are smaller than those 
expected from the relative errors in \ebv\ and \rv\ because these two quantities are usually anticorrelated in the likelihood CHORIZOS outputs.

\subsection{Experiment 4: New laws and variable \teff}

As previously mentioned, our fourth experiment is the equivalent to the second one with the new laws.
The comparison between the photometric temperatures derived from experiments 2 and 4 is shown in Figure~\ref{exp4}. From the graphical comparison it is clear that the results from
the fourth experiment are a significant improvement. Indeed, the mean difference between the photometric and spectroscopic temperatures has a mean of 2200~K (less than 1/3 of the
previous difference) and the normalized difference distribution now has a mean of $-0.78$ and a standard deviation of $1.41$ (compare to the ideal results of $0.0$ and $1.0$). These
results, though not perfect, are actually quite good. After all, typical spectroscopic determinations of \teff\ for O stars have uncertainties of 1000-2000 K. The results here have
typical random uncertainties of 2000~K for \teff~=~30\,000~K and 6000~K for \teff~=~45\,000~K. Hence, we can claim that {\it it is possible to photometrically measure the
effective temperature of an O star with [a] good accuracy (systematic biases comparable to random uncertainties, lower in most cases) and [b] good precision (random uncertainties 
only a factor of two higher than what is currently possible with spectroscopy)}. Spectroscopy provides better results by a factor of two and adds additional information on e.g. 
luminosity, metallicity, or $v\sin i$; so it is still preferred for detailed studies of individual objects. However, under the right conditions photometry yields acceptable \teff\ 
measurements for O stars (and even better ones for B stars, whose Balmer jump is more sensitive to temperature) with the advantage of its efficiency in terms of number of objects 
observed per unit of time.

We have covered a long distance since \citet{Hummetal88}, who entitled their paper ``Failure of continuum methods for determining the
effective temperature of hot stars'' and who started their abstract by stating: ``We demonstrate that for hot stars ($\teff > 30\,000$ K) methods based on the integrated continuum
flux are completely unreliable discriminators of the effective temperature''. Improvements on data, models, and techniques helped to fix the issues that plagued these authors.
As suspected by \citet{MaizSota08}, the last required step was an accurate extinction law.

\section{Discussion}

\subsection{How good are the new laws?}

In a nutshell, better than CCM but not perfect. A perfect extinction law should show symmetrical residuals with respect to zero in Figure~\ref{exp3} and also show a symmetrical
distribution with respect to zero in the vertical axis of Figure~\ref{exp4}. In principle, these issues could also be attributed to an incorrect spectral type-\teff\ scale or to
problems in the TLUSTY SEDs (more on that later), but given the large discrepancies observed in the first two experiments, it is more likely than an improvement in the extinction 
law could reduce the discrepancies even further. As stated elsewhere in this paper, {\it the final word on optical and NIR extinction laws will have to be provided by spectrophotometric
analyses}. A preliminary and limited study along that line is presented in the next two subsections.

\begin{figure*}
\centerline{\includegraphics[width=0.49\linewidth, bb=50 28 566 550]{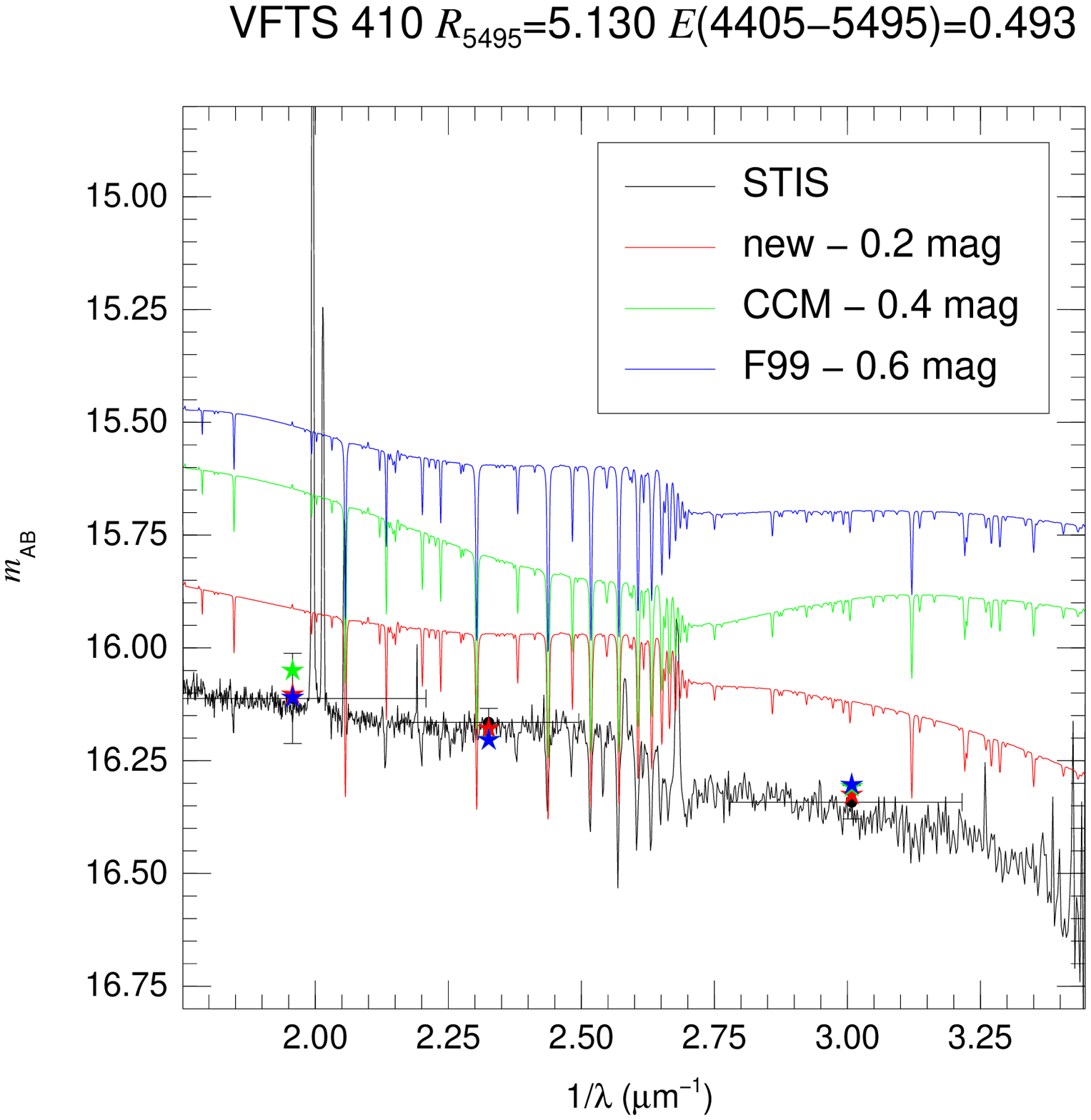} \
            \includegraphics[width=0.49\linewidth, bb=50 28 566 550]{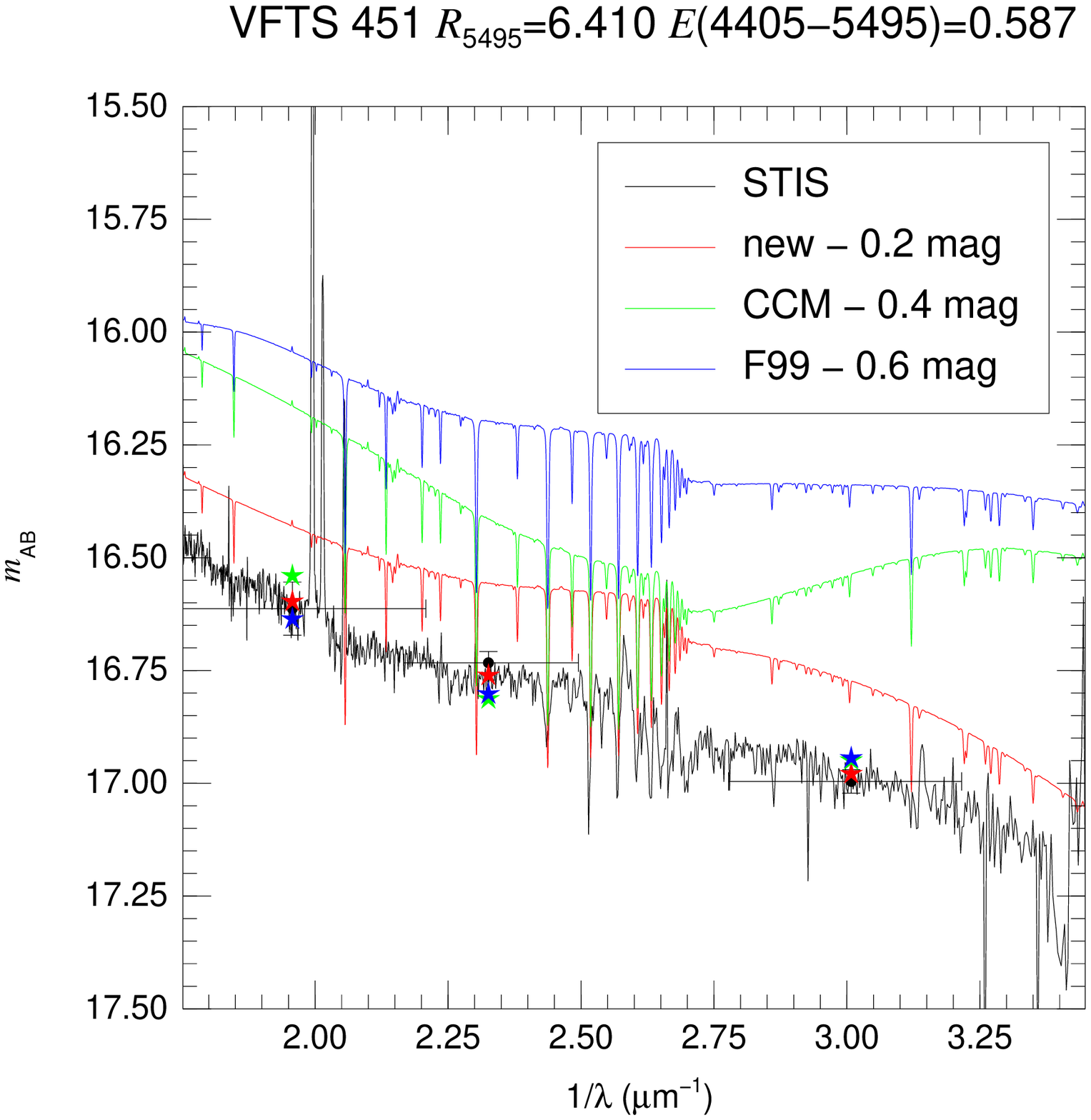}}
\caption{STIS spectrophotometry of VFTS 410 (left) and VFTS 451 (right) compared with their synthetic SEDs from experiment 1 (CCM, green), 3 (new, red), and alternate 1 (F99, blue)
shifted vertically for a better comparison. The error bars show the input F336W+F438W+F555W photometry and the colored star symbols the model photometry (with the same color coding
as the SEDs).}
\label{STIS_30Dor}
\end{figure*}

We should also ask ourselves why is it that the new laws are better than the CCM ones in the optical and NIR. The answer is threefold: better photometry, better technique, and better
sample. WFC3 photometry is better calibrated than Johnson's \citep{Maiz06a} and 2MASS has also allowed an all-sky calibration in the NIR. Spline interpolation is a more appropriate
technique than using a seventh-degree polynomial. But the largest difference is in the sample. CCM used only 29 stars, some with spectral types that have been updated since then, and their
\rv\ sample is heavily biased towards $\rv \sim 3.1$. Indeed, CCM had only one star with both $\rv> 5.0$ and $\ebv> 0.4$~mag, and that star (Herschel 36) turns out to have a NIR 
excess \citep{Ariaetal06}, to be a high-order multiple system \citep{Ariaetal10}, and to have its IUE spectra highly contaminated by the nearby Hourglass acting as a reflection nebula
(Ma{\'\i}z Apell\'aniz et al. in preparation): its use to derive an extinction law has to be considered very carefully, especially if it is the only representative of a category. On the other
hand, the sample in the new laws is almost three times larger and, more importantly, it covers the \rv\ range better (with the exception of $\rv < 3.0$).

\subsection{30 Doradus stars with STIS spectrophotometry}

Obtaining optical spectrophotometry of stars in 30 Doradus from the ground is difficult due to [a] crowding (which introduces additional stars in a wide slit) and [b] strong 
nebular contamination (which easily saturates the detector when obtaining a good S/N in the continuum). These two issues improve considerably when observing from space. Looking 
through the HST archive we found two stars in our sample, VFTS 410 and VFTS 451, with STIS G430L spectrophotometry 
\citep{Walbetal02a}. In Figure~\ref{STIS_30Dor} we compare the results of experiment 1 (F99 laws included) and experiment 3 with these data. Note that both stars have high values of \rv\ and 
above-average of \ebv, making them good choices to compare the discrepant part of the extinction laws.

The most obvious result is the confirmation that the functional form of the CCM laws provides the wrong wavelength behavior in the $UBV$ part of the spectrum. More specifically, we
confirm the existence of a significant $U$-band deficit for high values of \rv, as observed in the different slopes of the real and the synthetic spectra around $x\approx 2.9$~\mum1\
($\lambda\approx 3450$~\AA). The SEDs obtained with the new laws (and, to a lesser degree, the F99 ones\footnote{The high values of \chir\ for the F99 results arise largely from 
other photometric bands.}), on the other hand, follow the behavior with $\lambda$ reasonably well. A third interesting result is that there are no significant differences between the 
model and the real SED: in particular, the Balmer jump of these two O stars agrees with the one in the used TLUSTY models. 

In summary, the new laws survive this first limited test.

\subsection{Applying the new extinction laws to some Galactic cases}

\begin{figure*}
\centerline{\includegraphics[width=0.49\linewidth, bb=50 28 566 550]{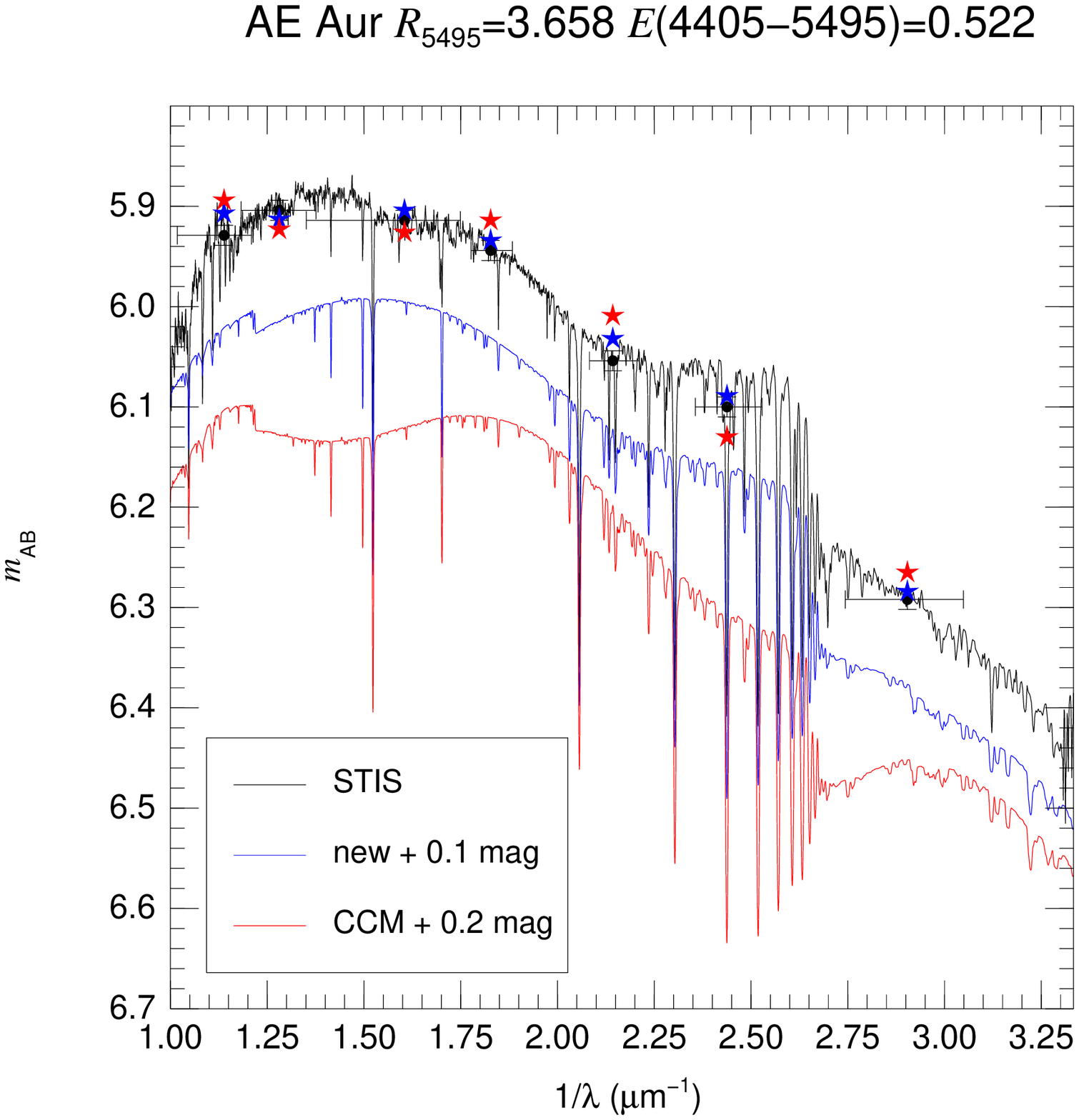} \
            \includegraphics[width=0.49\linewidth, bb=50 28 566 550]{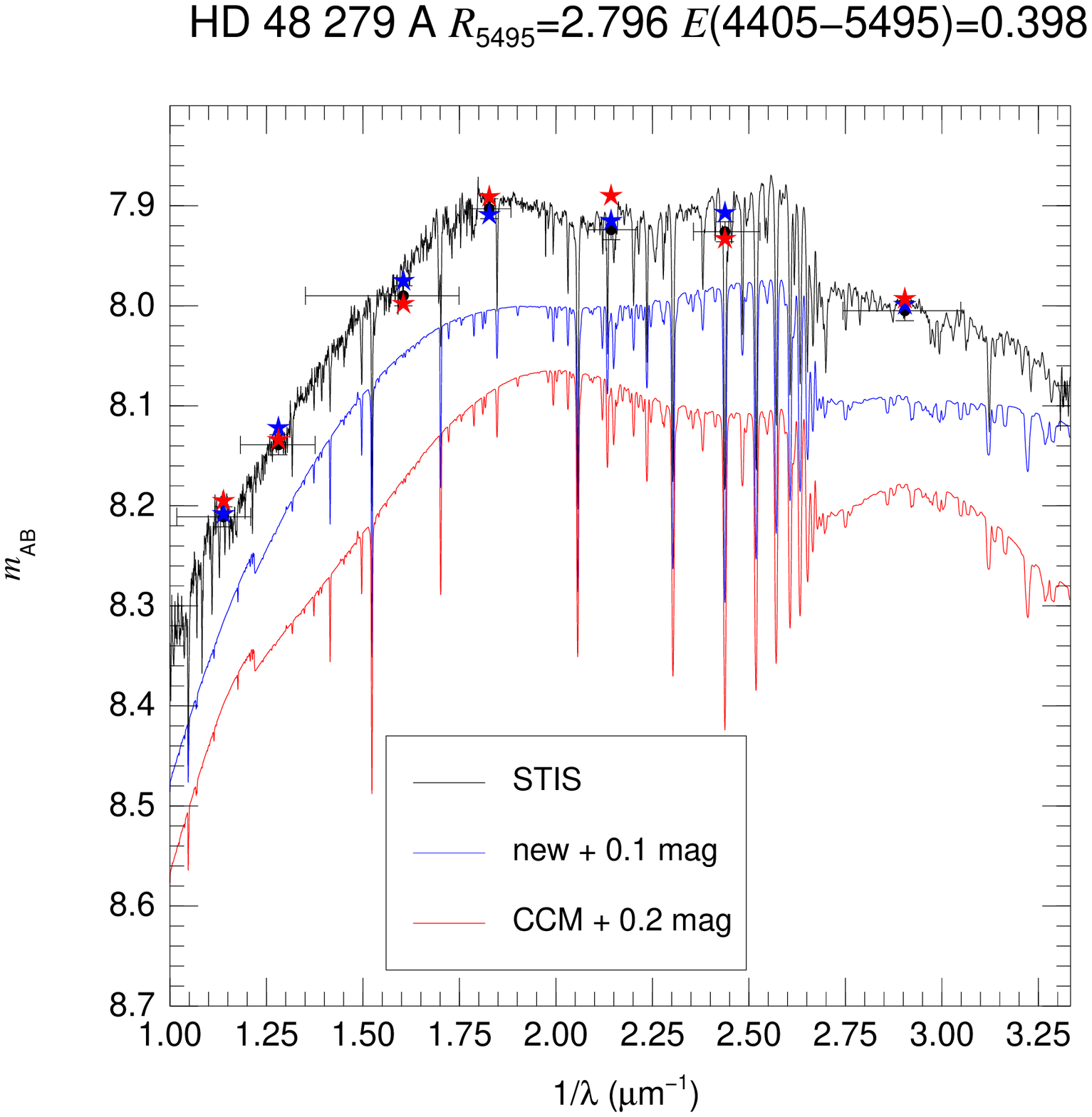}}
\caption{Comparison between the CCM and new extinction laws for two Galactic O stars with STIS spectrophotometry. Black is used for the input data, blue for the results with the new
extinction laws, and red for the results with CCM. The values for \rv\ and \ebv\ above are the ones derived with the new extinction laws.}
\label{STIS_Galactic}
\end{figure*}

A more stringent test of the new laws is their applicability to two regimes outside of the one in which they were derived: the Milky Way and higher values of \ebv\ (1.0-1.5~mag).
Regarding their applicability outside 30 Doradus, it is important to clarify an aspect that is sometimes overlooked by the non-specialist: the papers written over the past three decades
about the differences in the extinction law between the MW, the LMC, and the SMC (e.g. \citealt{Howa83,Prevetal84,Fitz85,GordClay98,Missetal99,MaizRubi12}) refer mostly to the UV
range. There is little work done on the possible differences in the optical and NIR (see \citealt{Tattetal13} for a recent example) 
and no definitive proof of significant differences between galaxies (as opposed to the UV, where clear
differences exist but which may be due to local environment conditions and not only to metallicity effects) for a given value of \rv. Therefore, there is no reason a priori to think
that the new laws are not applicable to the Milky Way\footnote{Note that since the beginning of this paper we have applied the inverse argument: that CCM laws are potentially
applicable to 30 Doradus even though they were derived for the MW.}.

Two moderately extinguished Galactic O stars (AE~Aur and HD~48\,279~A with even better STIS coverage (G430L and G750L) than for the two 30 Doradus stars in the previous subsection 
are present in the HST archive thanks to the Next Generation Spectral Library \citep{Gregetal04}. The spectra were reprocessed and calibrated in flux using Tycho-2 photometry
\citep{Maiz05b,Maiz06a,Maiz07a}. We derived synthetic photometry from the spectrophotometry, estimated their \teff\ from their spectral types \citep{Sotaetal11a}, and applied CHORIZOS
in a manner equivalent to experiments 1 (CCM) and 3 (new)\footnote{With one difference: we fix the luminosity class and leave distance as a free parameter.}. Results are shown in 
Figure~\ref{STIS_Galactic}.

The most obvious conclusion from Figure~\ref{STIS_Galactic} is that the new extinction laws reproduce the detailed behavior of the extinction law in the optical range better than the 
CCM ones: once again, splines beat a seventh degree polynomial. In particular, the behavior around the \cite{Whit58} 2.2~\mum1\ knee and the $R$ band ($\sim$1.6~\mum1) 
is better reproduced. Another 
positive conclusion is that the Balmer jump for these later-type O stars is also well reproduced by the TLUSTY models, another indication that it is not the source of discrepancies for 
\teff\ in experiments 2 and 4. There are also two not-so-positive conclusions. The first one is that the detailed wavelength behavior of HD~48\,279~A fit with the new laws is not as 
good as the one for AE~Aur (though it is still a slight improvement over the CCM fit). This is not a surprising conclusion because the value of \rv\ for HD~48\,279~A is outside the
range measured in 30 Doradus while the one for AE~Aur is inside: extrapolated laws are more uncertain than interpolated ones. The second one is the realization that TLUSTY
models (at least those of \citealt{LanzHube03,LanzHube07}, note that in this case we are using the grid with MW metallicity, not the LMC one) do not treat the Paschen jump correctly, 
apparently because they do not include the higher-order transitions. Hence, one should be careful when analyzing $iz$ photometry with TLUSTY models. 

\begin{figure*}
\centerline{\includegraphics[width=0.49\linewidth, bb=50 28 566 550]{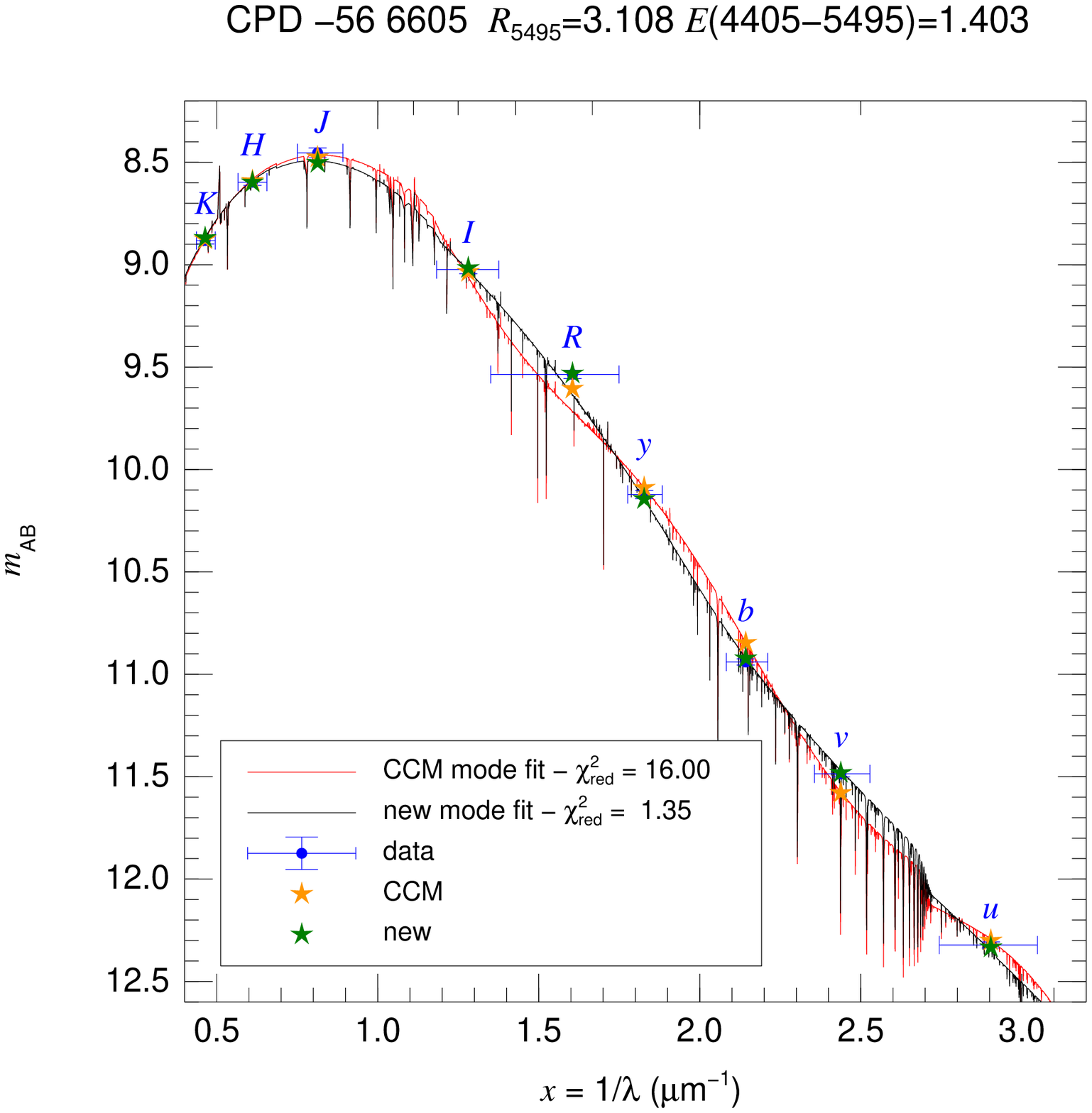} \
            \includegraphics[width=0.49\linewidth, bb=50 28 566 550]{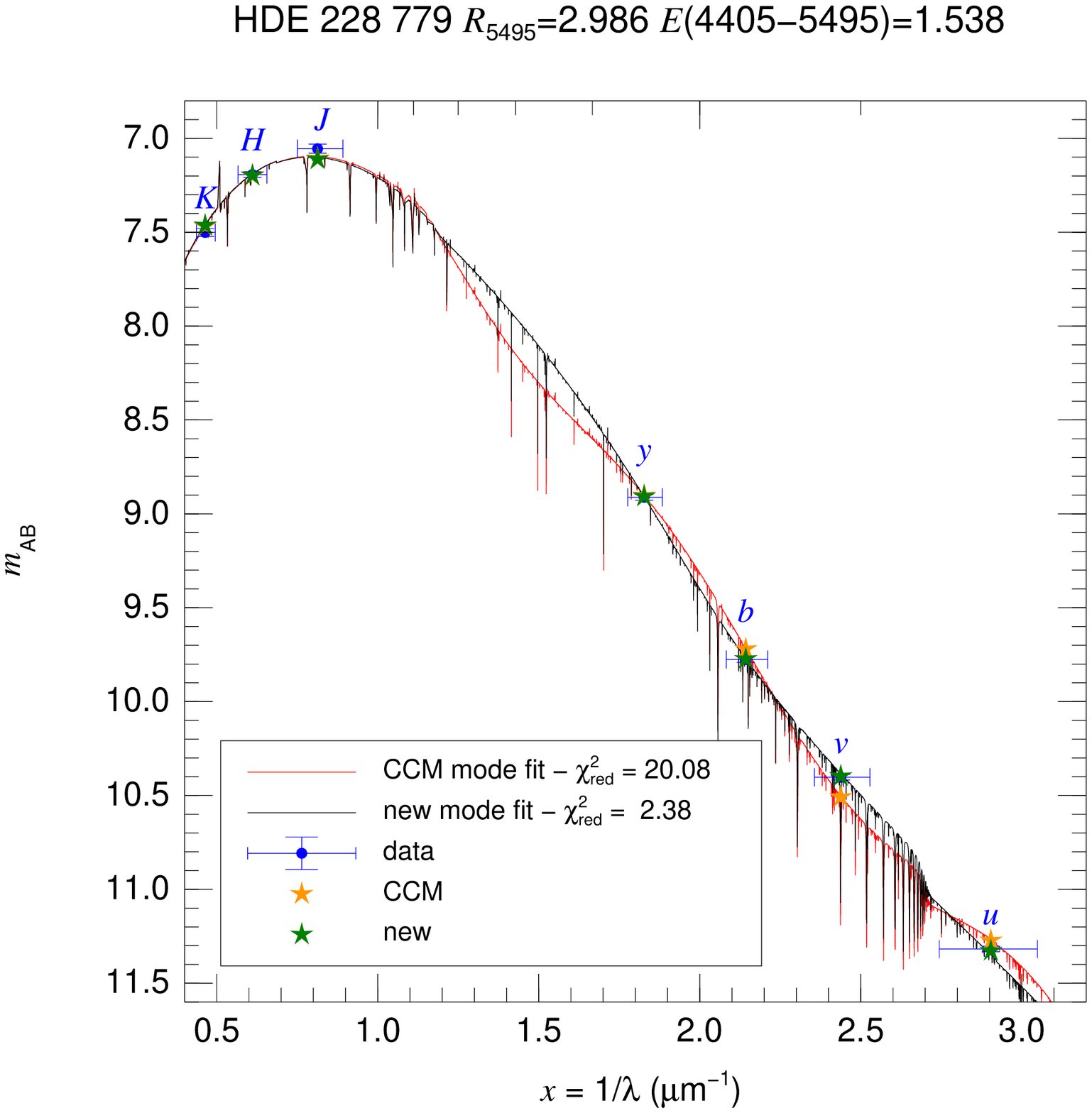}}
\caption{Comparison between the CCM and new extinction laws for two extinguished Galactic O stars with 2MASS and Str\"omgren photometry (the first one also with Cousins $RI$)
processed with CHORIZOS. The blue points with error bars (horizontal for indicative filter extent, vertical for uncertainty) show the input photometry used for the fit. Red and orange are 
used for the SEDs and synthetic photometry derived with the CCM extinction laws while black and green are used for the same quantities derived with the new extinction laws. The values for 
\rv\ and \ebv\ above are the ones derived with the new extinction laws.}
\label{Stromgren_Galactic}
\end{figure*}

What about higher extinctions? We searched the Galactic O-Star Catalog \citep{Maizetal04b,Sotaetal08} for O stars with $\ebv\sim 1.5$~mag and good-quality Str\"omgren $uvby$ and 2MASS 
$JHK_{\rm s}$ photometry\footnote{The reason for preferring Str\"omgren to Johnson photometry is that the use of two filters ($vb$) in the $B$-band region makes the first system
more sensitive to the behavior of the extinction law across the \citet{Whit58} 2.2~\mum1\ knee.}. We found two stars, CPD~$-$56~6605 and HDE~228\,779, that met the requirements; the
first one also has Cousins $RI$ photometry available. Neither of the two stars is included in the first two papers \citep{Sotaetal11a,Sotaetal14} of the Galactic
O-Star Spectroscopic Survey \citep{Maizetal11}, but the project has already obtained their spectra and classified them as O9.7~Iab and O9~Iab, respectively, hence allowing us to
derive their \teff. With that information, we performed the CHORIZOS experiments in analogy to the ones for AE~Aur and HD~48\,279~A. Results are shown in 
Figure~\ref{Stromgren_Galactic}.

Once again, the new laws beat the CCM ones by a large margin. The CCM values for \chir\ of 16.00 and 20.08 for the two stars are reduced to 1.35, and 2.38, respectively. Also, the 
detailed wavelength behavior of the new SED does not show the wiggles present in the CCM one. Therefore, at least for these two cases the new laws provide a good fit to heavily 
extinguished O stars with standard values of \rv\ (3.0-3.1)\footnote{The difference between the two extinction law families is, in general, small for quantities that 
depend on the value of $F_\lambda$ (e.g. magnitudes and $A_V$), larger for those that depend on its first derivative (e.g. colors), and even larger for those that depend on its second
derivative (e.g. Str\"omgren $m_1$ and $c_1$ indices).}.

\section{Guidelines and future work}

When applying the extinction laws in this paper, we recommend following these guidelines:

\begin{itemize}
 \item In general, passband effects (differences between the extinction measured at the central wavelength of a filter and the extinction integrated
       over the whole passband) can be significant (\citealt{Maiz13b} and references therein), especially for samples with large extinctions and 
       differences in \teff. Do not apply simple linear extinction corrections (e.g. $Q$-like parameters) in these cases. Instead, integrate over 
       the whole passband using the routine in Table~\ref{IDL} or its equivalent.
 \item In the optical, the extinction laws have been tested on a limited sample up to $\ebv\approx 1.5$~mag. Future studies may improve the results here,
       though it would be surprising if an optical extinction with an overall dependence with wavelength very different from the ones here or in CCM 
       were discovered.
 \item The NIR law used here is the power law of \cite{RiekLebo85}. A large number of works since then (e.g. \citealt{Mooretal05,FitzMass09,Fritetal11}) have found 
       different values of the power law exponent and possible changes between sightlines so, strictly speaking, the laws here should not be applicable. However, 
       for low and intermediate reddenings $\ebv \le 2.0$~mag, NIR extinction corrections should be small enough to yield acceptable differences. In other words, in 
       the NIR apply these laws to optically visible OB stars, not to targets in the Galactic Center. 
 \item It is well known that the extinction law in the MIR and FIR is not a power law. Do not apply the results here in that range (that is the reason why 
       the routine in Table~\ref{IDL} does not accept values below $x = 0.3$~\mum1).
 \item UV extinction is a different and difficult issue\footnote{For example, \citet{FitzMass07} find that there is no strong correlation between the UV and the IR 
       Galactic extinctions, in opposition to what CCM found. More recently, \citet{PeekSchi13} have found that at high Galactic latitudes UV extinction is anomalous 
       and suggested a connection with an increase in the amount of very small silicate grains.} that is not measured in this paper even though a functional form is provided. 
       More specifically, the routine in Table~\ref{IDL} may work for low values of \rv\ (as CCM does) but is guaranteed to fail for high values of \rv. The 
       reason is that the jump seen for \rv\ = 5.0 and 7.0 around $x = 3.9$~\mum1\ is not physical but a product of tying up the results of this paper in 
       the optical with those of CCM in the UV. 
\end{itemize}

Our future work will develop along the following lines:

\begin{itemize}
 \item We will analyze the spatial distribution of dust in 30 Doradus in a subsequent paper of the VFTS series. In particular, we will study the dependence of
       \rv\ with the environment.
 \item We will apply the optical and NIR extinction laws in this paper to a number of existing datasets. In particular, we will use them to measure the amount and type
       of extinction in the GOSSS stars \citep{Maizetal11}. 
 \item We will use spectrophotometry to check the detailed behavior in $\lambda$ of the extinction laws.
 \item We will obtain data for stars with $\ebv = 1.5-3.0$~mag to test the relationship between \rv\ and the NIR slope.
\end{itemize}

\begin{acknowledgements}
This article is based on observations at the European Southern Observatory Very Large Telescope in programme 182.D-0222 and on
observations made with the NASA/ESA Hubble Space Telescope (HST) associated with GO program 11\,360 and obtained at the Space 
Telescope Science Institute, which is operated by the Association of Universities for Research in Astronomy, Inc., under NASA 
contract NAS 5-26\,555. We would like to thank Max Mutchler for helping with the processing of the WFC3 images. J.M.A. acknowledges
support from [a] the Spanish Government Ministerio de Econom{\'\i}a y Competitividad (MINECO) through grants AYA2010-15\,081 and AYA2010-17\,631,
[b] the Consejer{\'\i}a de Educaci{\'o}n of the Junta de Andaluc{\'\i}a through grant P08-TIC-4075, and [c] the George P. and 
Cynthia Woods Mitchell Institute for Fundamental Physics and Astronomy. He is also grateful to the Department of Physics and 
Astronomy at Texas A\&M University for their hospitality during some of the time this work was carried out.
R.H.B. acknowledges support from DIULS and FONDECYT Project 1\,120\,668.
A.H. and S.S.-D. acknowledge funding by [a] the Spanish Government Ministerio de Econom{\'\i}a y Competitividad (MINECO) through 
grants AYA2010-21\,697-C05-04, AYA2012-39\,364-C02-01, and Severo Ochoa SEV-2011-0187 and [b] the Canary Islands Government under grant PID2\,010\,119.
\end{acknowledgements}

\begin{figure}
\includegraphics[width=\linewidth]{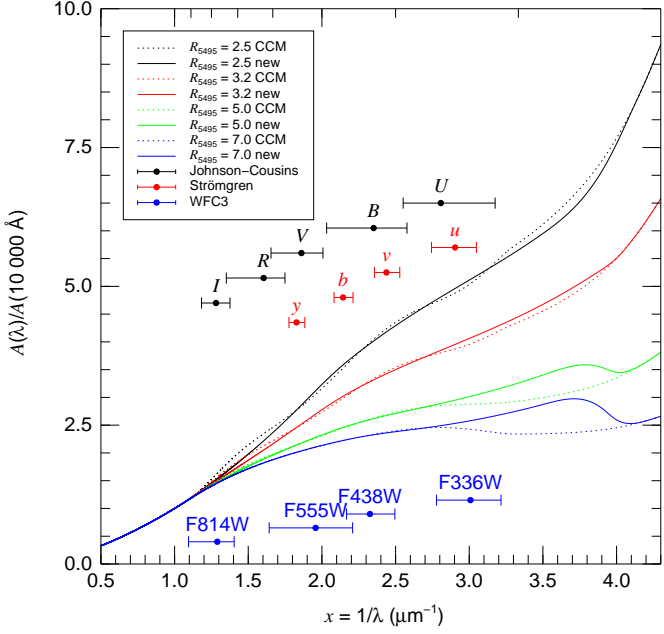}
\caption{CCM and new extinction laws for four values of \rv\ (2.5, 3.2, 5.0, and 7.0). Extinction is normalized to the value at 10\,000 \AA\ in each 
case to emphasize that the extinction laws are the same for longer wavelengths and to better visualize the differences in the optical and NUV 
ranges. The extent of some filters in three common systems (Johnson-Cousins, Str\"omgren, and WFC3) is shown.}
\label{extlaws1}
\end{figure}

\begin{figure}
\includegraphics[width=\linewidth]{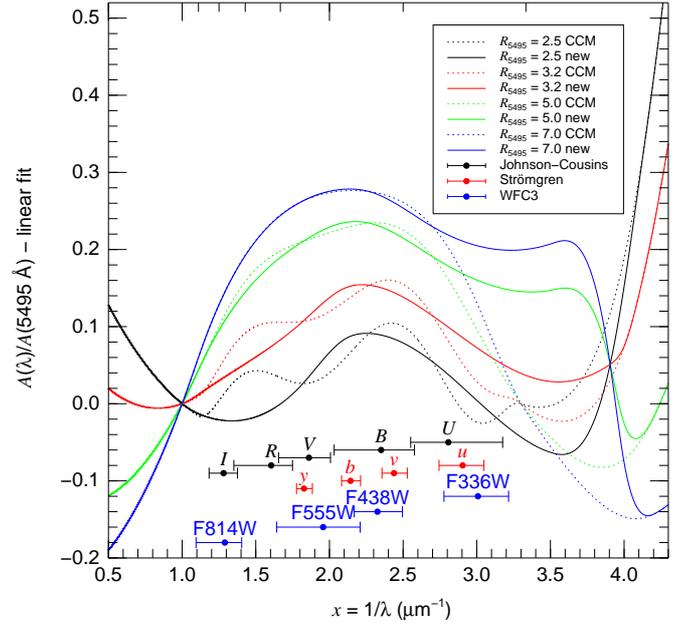}
\caption{As Fig.~\ref{extlaws1} but with a different normalization to emphasize the differences between extinction laws. In each case we have 
subtracted a linear fit $A(\lambda)/A(5495 \AA) = a(\rv) + b(\rv)x$, with $a(\rv)$ and $b(\rv)$ calculated so that the CCM law for that \rv\ is 0.0 
at $x=1.1$~\mum1\ and at $x=3.3$~\mum1, the limits for the optical range in CCM.}
\label{extlaws2}
\end{figure}

\bibliographystyle{aa}
\bibliography{general}

\appendix

\section{The new extinction laws}
\label{newlaws}

Based on the issues discussed in \citet{Maiz13b} and the results of experiments 1 and 2, we decided to attempt the calculation of a new
family of extinction laws. Ideally, to complete such a task one would use high-quality spectrophotometry from the NIR to the UV of a diverse
collection of sources in different environments and with different degrees of extinction. Since such dataset is not available, in this paper we
concentrate on only some of the problems discussed in \citet{Maiz13b}. More specifically, we will ignore extinction in the UV (except for the
region closest to the optical) and for the NIR we will simply use the CCM laws (which, in turn, used \citealt{RiekLebo85})\footnote{We have verified
that our values of $E(J-H)$ and $E(H-K)$ are compatible with the \citet{RiekLebo85} value for the NIR exponent and with the 
TLUSTY SEDs yielding the correct intrinsic colors for OB stars with LMC metallicity. Note, however, that given the low values of the NIR extinction 
of the stars in our sample, there is not enough information to discard alternatives.}. In other words, we will concentrate on the optical region since that
is the critical component for the determination of \teff\ for OB stars. Ignoring the UV will not matter to a non-specialist interested only in
eliminating the extinction from his/her optical data. Ignoring the NIR may matter if the exponent there is significantly different from the CCM one
but only if extinction is very large and even then it may only apply to the total extinction correction, not to the determination of \teff\ from
the photometry.

The immediate goals of the new family of extinction laws are:

\begin{enumerate}
  \item To maintain the overall properties of the CCM laws that have made them so successful: a single-parameter, easy-to-calculate family 
        that covers a large wavelength range; and their overall shape as a function of wavelength (including the functional form in Eqn.~\ref{rvdependence}).
  \item To eliminate the F336W ($U$-band like) excesses detected in experiment 1, which lead to the temperature biases in experiment 2.
  \item To at least alleviate the wiggles induced by the seventh-degree polynomial used by CCM in the optical range and make the new family more similar
        to the shape derived by \cite{Whit58} with spectrophotometry (two straight lines with a knee at $x = 2.2$~\mum1).
\end{enumerate}

To achieve those goals, we use the following strategy:

\begin{enumerate}
  \item Instead of a seventh-degree polynomial, we [a] select a series of points in $x$ in the optical range, [b] use the values of the CCM laws
        at these points (with corrections in some cases), and [c] apply a spline interpolation between these points. Note that a spline interpolation was
        already used by \citet{Fitz99}.
  \item We expand the optical range from $x$ = 1.1-3.3 \mum1\ to 1.0-4.2 \mum1\ in order to avoid discontinuities and/or knees near the edges of the 
        ranges\footnote{Note that such knees exist for low values of \rv\ in the CCM laws, see the \rv\ = 2.5 case in Fig.~\ref{extlaws2}.}.
  \item The first two points selected are $x$ = 1.81984 \mum1\ and $x$~=~2.27015 \mum1, which correspond to 5495 \AA\ and 
        4405~\AA. The choice is determined by the need to maintain the values of \rv\ for a given extinction law. At these values no correction is
        applied to the CCM laws.
  \item A third point is added between $x$ = 1.81984 \mum1\ and $x$~=~1.0~\mum1\ to minimize the wiggles visible for low and intermediate values of 
        \rv\ in CCM and thus make the extinction laws more similar to that of \citet{Whit58}. After trying different choices, we select 
        $x=1.15$ \mum1\ as the one that produces the smoothest results. Note that the values of the extinction law at exactly these three points are 
        the CCM ones: the changes affect the points in between due to the use of a spline interpolation instead of a seventh-degree polynomial.
  \item A fourth point is added between 5495 \AA\ and 4405 \AA\ to maintain the \citet{Whit58} knee at its original location near $x$~=~2.2~\mum1\
        (as it can be seen in Fig.~\ref{extlaws1}, CCM moved the knee towards higher values, i.e. shorter wavelengths, in most cases while the new 
        laws put it back between 2.15~\mum1\ and 2.25~\mum1\ for most values of \rv). By trial and error we selected $x$~=~2.1~\mum1\
        and applied a correction to the \al/$A(5495)$ CCM values there of $-0.011 + 0.091/\rv$.
  \item Five final points are added between $x$ = 2.27015 \mum1\ and $x$ = 4.2 \mum1. Different combinations were tried with the general goals of 
        [a] keeping smooth profiles, [b] correcting the overall F336W excesses found in experiment 1 as a function of \ebv, and [c] doing the same as 
        a function of \rv. The final result leads to the points being located at $x$~=~2.7~\mum1, 3.5~\mum1, 3.9~\mum1, 4.0~\mum1, and 4.1~\mum1. The 
        corrections to \al/$A(5495)$ in these points are 0, $0.442 - 1.256/\rv$, $0.341 - 1.021/\rv$, $0.130 - 0.416/\rv$, and $0.020 - 0.064/\rv$, 
        respectively. Note that the optical region in CCM ends at $x$~=~3.3~\mum1: for higher values of $x$ the correction is applied to the CCM UV functional form.
\end{enumerate}

Four examples of the new extinction laws are shown in Figs.~\ref{extlaws1}~and~\ref{extlaws2}. An IDL function to obtain the new extinction
laws is provided in Table~\ref{IDL}. The validity of the new extinction laws is tested in experiments 3 and 4 i.e. the ones used to iteratively determine them.

\begin{table*}
\caption{IDL coding of the extinction laws in this paper.}
{\small
\begin{verbatim}
FUNCTION ALA5495, lambda, R5495=r5495
; This function gives A_lambda/A_5495 for the range 1000 Angstroms - 33 333 Angstroms
;  (3.333 3 microns) according to the Ma\'{\i}z Apell\'aniz et al. (2014) extinction laws.                                                            
; Positional parameters:                                                       
; lambda:   Wavelength in Angstroms (single value or array).                         
; Keyword parameters:                                                          
; R5495:   R_5495 value. By default, it is 3.1.                                
IF KEYWORD_SET(R5495) EQ 0 THEN r5495 = 3.1
x   = 10000D/lambda
n   = N_ELEMENTS(x)
IF MIN(x) LT 0.3 OR MAX(x) GT 10.0 THEN STOP, 'Wavelength not implemented'
; Infrared
ai  =   0.574*x^1.61
bi  = - 0.527*x^1.61
; Optical
x1  = [1.0]
xi1 = x1[0]
x2  = [1.15,1.81984,2.1,2.27015,2.7]
x3  = [3.5 ,3.9    ,4.0,4.1    ,4.2]
xi3 = x3[N_ELEMENTS(x3)-1]
a1v =   0.574      *x1^1.61
a1d =   0.574*1.61*xi1^0.61
b1v = - 0.527      *x1^1.61
b1d = - 0.527*1.61*xi1^0.61
a2v =  1 + 0.17699*(x2-1.82)   - 0.50447*(x2-1.82)^2 - 0.02427*(x2-1.82)^3 + 0.72085*(x2-1.82)^4 $
         + 0.01979*(x2-1.82)^5 - 0.77530*(x2-1.82)^6 + 0.32999*(x2-1.82)^7 + [0.0,0.0,-0.011,0.0,0.0]
b2v =      1.41338*(x2-1.82)   + 2.28305*(x2-1.82)^2 + 1.07233*(x2-1.82)^3 - 5.38434*(x2-1.82)^4 $
         - 0.62251*(x2-1.82)^5 + 5.30260*(x2-1.82)^6 - 2.09002*(x2-1.82)^7 + [0.0,0.0,+0.091,0.0,0.0]
a3v =   1.752 - 0.316*x3 - 0.104/             (( x3-4.67)*( x3-4.67) + 0.341) + [0.442,0.341,0.130,0.020,0.000] 
a3d =         - 0.316    + 0.104*2*(xi3-4.67)/((xi3-4.67)*(xi3-4.67) + 0.341)^2   
b3v = - 3.090 + 1.825*x3 + 1.206/             (( x3-4.62)*( x3-4.62) + 0.263) - [1.256,1.021,0.416,0.064,0.000] 
b3d =         + 1.825    - 1.206*2*(xi3-4.62)/((xi3-4.62)*(xi3-4.62) + 0.263)^2
as  =           SPL_INIT([x1,x2,x3], [a1v,a2v,a3v], YP0=a1d, YPN_1=a3d)
bs  =           SPL_INIT([x1,x2,x3], [b1v,b2v,b3v], YP0=b1d, YPN_1=b3d)
av  = REVERSE(SPL_INTERP([x1,x2,x3], [a1v,a2v,a3v], as, REVERSE(x)))
bv  = REVERSE(SPL_INTERP([x1,x2,x3], [b1v,b2v,b3v], bs, REVERSE(x)))
; Ultraviolet
y   = x - 5.9
fa  = REPLICATE(0.0D,n) + (- 0.04473*y^2 - 0.009779*y^3)*(x LE 8.0 AND x GE 5.9)
fb  = REPLICATE(0.0D,n) + (   0.2130*y^2 +   0.1207*y^3)*(x LE 8.0 AND x GE 5.9)
au  =   1.752 - 0.316*x - 0.104/((x-4.67)*(x-4.67) + 0.341) + fa
bu  = - 3.090 + 1.825*x + 1.206/((x-4.62)*(x-4.62) + 0.263) + fb
; Far ultraviolet
y   = x - 8.0
af  = -  1.073 - 0.628*y + 0.137*y^2 - 0.070*y^3
bf  =   13.670 + 4.257*y - 0.420*y^2 + 0.374*y^3
; Final result
a   = ai*(x LT xi1) + av*(x GE xi1 AND x LT xi3) + au*(x GE xi3 AND x LT 8.0) + af*(x GE 8.0)
b   = bi*(x LT xi1) + bv*(x GE xi1 AND x LT xi3) + bu*(x GE xi3 AND x LT 8.0) + bf*(x GE 8.0)
RETURN, a + b/r5495
END
\end{verbatim}
}
\label{IDL}
\end{table*}

\section{CHORIZOS and SED models}
\label{CHORIZOSgrids}

The code CHORIZOS was presented in \citet{Maiz04c} as a $\chi^2$ Code for Parameterized Modeling and Characterization of Photometry and 
Spectrophotometry. In subsequent versions, it evolved to become a complete bayesian code that matches photometry and spectrophotometry to 
spectral energy distribution (SED) models in up to six dimensions. Some examples of its applications can be seen in 
\citet{Maizetal04d,Maizetal07,Neguetal06,Ubedetal07a}. The last public version of CHORIZOS, v. 2.1.4, was released in July 2007. Since then, the
first author of this paper has been working on versions 3.x, which, among many changes, allow for the use of magnitudes (instead of colors) as
fitting quantities and the use of distance as an additional parameter. Problems with the code speed and memory usage did not allow these versions of
CHORIZOS to become public (even though the code itself worked for restricted cases). The largest problems have now been solved and a public application
with the 3.x version of the code will be publicly available soon. 

The use of magnitudes and distances described above leads to the possibility of a new type of stellar SED grids: instead of using \teff\
and $\log g$ as the two parameters, one can substitute $\log g$ by luminosity or an equivalent parameter \citep{Maiz13a}. As an intermediate step, one 
needs to use evolutionary tracks or isochrones that assign the correct value of $\log g$ to a given luminosity. We have developed a class of such grids 
with the following characteristics:

\begin{enumerate}
  \item There are three separate grids corresponding to the Milky Way, LMC (the ones used in this paper), and SMC metallicities.
  \item The luminosity-type parameter is called (photometric) luminosity class and is analogous to the spectroscopic equivalent. To maintain the
        equivalence as close as possible, its value ranges from 0.0 (hypergiants) to 5.5 (ZAMS).
  \item The grids use Geneva evolutionary tracks for high-mass stars and Padova ones for intermediate- and low-mass stars. For the objects in this
        paper, the relevant tracks are those of \citet{Schaetal92,Schaetal93a}. Note that the use of tracks with rotation would not introduce 
        significant changes in the results of this paper because the purpose of the tracks 
        for O stars are to [a] establish the total range in luminosities and [b] determine the gravity for a given temperature and luminosity. The range 
        in luminosities changes little with the introduction of rotation and the possible changes in gravity at a given grid point can be of the order 
        of 0.2 dex, which leads to an insignificant effect in the optical colors of O stars\footnote{There are some cases (very high mass, extreme $v\sin i$)
        where rotation does matter but they are not relevant to this paper.}. 
  \item Different SEDs are used as a function of temperature and gravity (or luminosity). For the objects in this paper, the relevant SEDs are the
        two TLUSTY grids of \citet{LanzHube03,LanzHube07}.
  \item It is possible to specify a total of five parameters in a given grid: \teff, luminosity class, \ebv, \rv, and distance. Note, however, that
        for most practical applications it is only possible to leave four of these parameters free.
  \item For the experiments in this paper we have calculated independent grids with the CCM, F99, and new extinction laws.
\end{enumerate}

\begin{figure*}
\includegraphics[width=\linewidth]{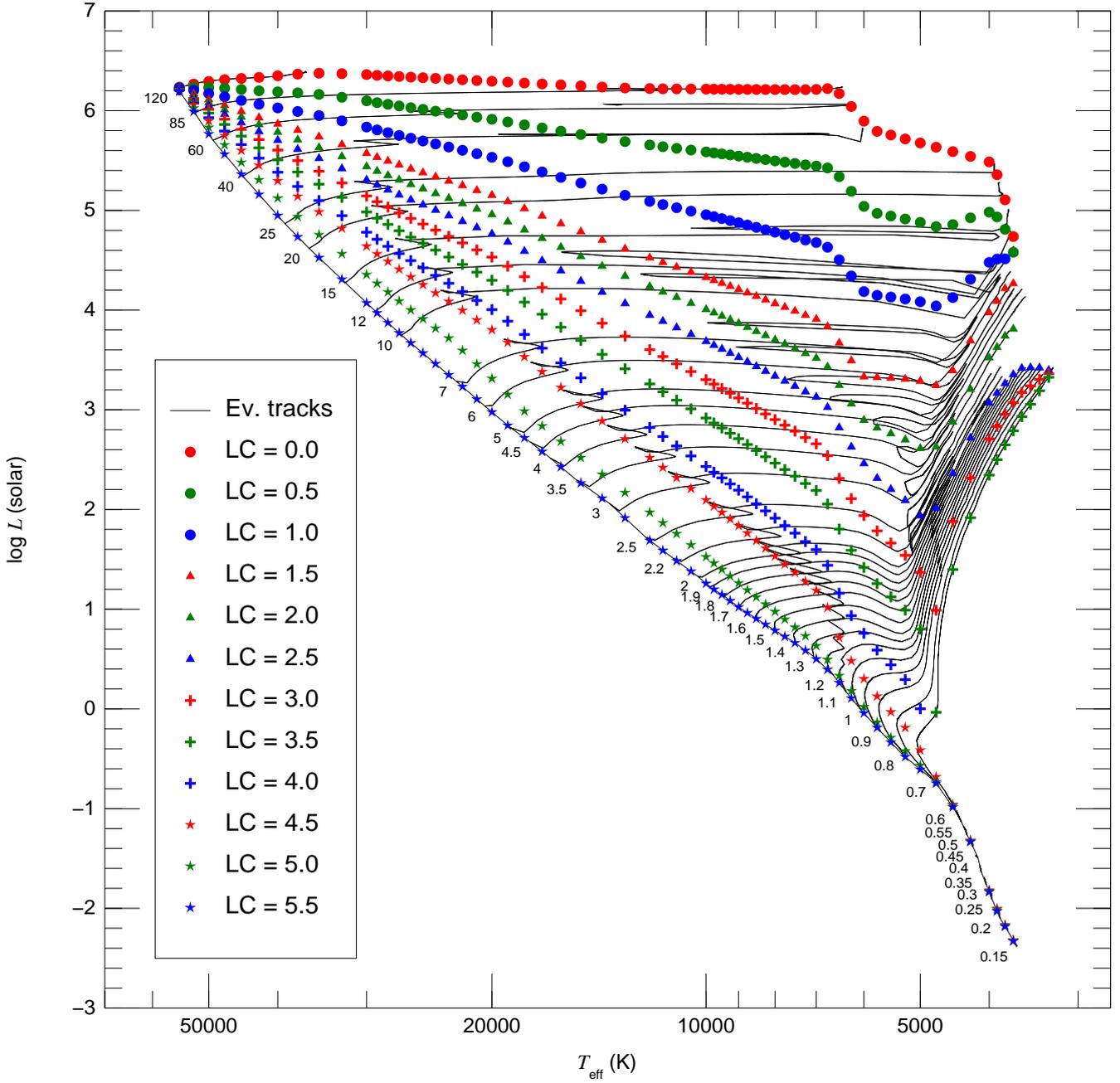}
\caption{The \teff-luminosity class distance-calibrated SED family for the LMC developed for CHORIZOS. The black lines are the Geneva/Padova 
evolutionary tracks between 0.15 M$_\odot$ and 120 M$_\odot$ (a label at the beginning of the track shows the initial mass). Different symbols 
are used for the luminosity types 0.0, 0.5\ldots 5.5. Note that luminosity types are defined at 0.1 intervals but only those at 0.5 intervals are 
shown for clarity.}
\label{LMCgrid}
\end{figure*}

Figure~\ref{LMCgrid} shows the LMC grid used in this paper.

\section{Extinction along a sightline with more than one type of dust}

In this paper we make no attempt to disentangle the contributions to the extinction in 30 Doradus among its three possible components: Milky Way (MW),
Large Magellanic Cloud (LMC), and internal (30 Dor). The reason for not attempting to do so is the impossibility of doing it with the available data (each
component can contribute while being spatially variable, see \citealt{vanLetal13}). However, that does not affect the validity of our results due to a feature of
the extinction law families of CCM and this paper. The extinction laws are written as:

\begin{equation}
\al/A(5495) = a(\lambda) + b(\lambda)/\rv,
\label{rvdependence}
\end{equation}

\noindent where $a(\lambda)$ and $b(\lambda)$ are defined by different functional forms in different wavelength ranges (see e.g. Table~\ref{IDL}). A SED with
an original form of $I_0(\lambda)$ is extinguished to $I(\lambda)$ according to:

\begin{equation}
\begin{array}{rcl}
I(\lambda) & = & I_0(\lambda) 10^{-0.4A_\lambda}                                               \\
           & = & I_0(\lambda) 10^{-0.4E(4405-5495)R_{5495}(a(\lambda) + b(\lambda)/R_{5495})}. \\
\end{array}
\label{exteq1}
\end{equation}

Now, suppose that along the sightline to a star there are $N$ clouds, each one of them with a color excess $\ebv_i$ and type of extinction $\rv_{,i}$. The total 
extinction will be:

\begin{equation}
I(\lambda) = I_0(\lambda) \prod_{i=1}^N 10^{-0.4E(4405-5495)_i R_{5495,i}(a(\lambda) + b(\lambda)/R_{5495,i})}. \\
\label{exteq2}
\end{equation}

It is easy to show that Eqns.~\ref{exteq1}~and~\ref{exteq2} are equivalent if one defines:

\begin{equation}
E(4405-5495) = \sum_{i=1}^N E(4405-5495)_i
\end{equation}

\begin{equation}
R_{5495} = \frac{\sum_{i=1}^N E(4405-5495)_iR_{5495,i}}{\sum_{i=1}^N E(4405-5495)_i},
\end{equation}

\noindent i.e. if each individual extinction law belongs to the same family, then it will represent the combined effect of the $N$ clouds. The total color excess is 
simply the sum of the individual ones and the type of extinction is the sum of the individual types (characterized by their \rv\ values) weighted by the individual color excesses.

\end{document}